\documentclass[useAMS,usenatbib,onecolumn]{mn2e}
\usepackage{graphicx}
   \title{Magnetic White Dwarf Stars in the Sloan Digital Sky Survey}

   \author[Kepler et al.]{S. O. Kepler$^{1}$\thanks{kepler@if.ufrgs.br},
		I. Pelisoli$^{1}$,
                S. Jordan$^2$,
		S. J. Kleinman$^{3}$,
\newauthor
		D. Koester$^{4}$,
                B. K\"ulebi$^{5,6}$,
		V. Pe\c{c}anha$^{1}$,
		B. G. Castanheira$^{7,8}$,
		A. Nitta$^{3}$,
\newauthor
                J.E.S. Costa$^{1}$,
		D. E. Winget$^{8}$,
                A. Kanaan$^9$,
                and
                L. Fraga$^{10}$\\
$^{1}$Instituto de F\'{\i}sica, Universidade Federal do Rio Grande do Sul,
              91501-900  Porto-Alegre, RS, Brazil\\
$^2$Astronomisches Rechen-Institut, Zentrum f\"ur Astronomie der Universit\"at Heidelberg, M\"onchhofstr. 12-14, \\ D-69120 Heidelberg, Germany\\
$^{3}$Gemini Observatory, Hilo, Hawaii, 96720, USA\\
$^{4}$Institut f\"ur Theoretische Physik und Astrophysik, Universit\"at Kiel, 24098 Kiel, Germany\\
$^{5}$Institut de Ci\`encies de L'Espai, Universitat Aut\`onoma de Barcelona, Barcelona, Spain\\
$^{6}$Institute for Space Studies of Catalonia, c/Gran Capit\`a 2-4, Edif. Nexus 104, 08034 Barcelona, Spain\\
$^{7}$ Institut f\"ur Astronomie, Wien, Austria\\
$^8$McDonald Observatory and Department of Astronomy, University of Texas, Austin, TX, 78712, USA\\
$^9$ Universidade Federal de Santa Catarina, Florian\'opolis, SC, Brazil\\
$^{10}$ Soar Telescope, La Serena, Chile
}
\begin{document}
\date{Accepted  Received }

\pagerange{\pageref{firstpage}--\pageref{lastpage}} \pubyear{2012}
   \maketitle

\label{firstpage}
  \begin{abstract}

To obtain better statistics on the occurrence of magnetism among white dwarfs, we 
searched  the spectra of the hydrogen atmosphere white dwarf stars (DAs) in the Data
Release~7 of the Sloan Digital Sky Survey (SDSS) for Zeeman splittings and estimated the magnetic fields.
 We found 521 DAs with detectable Zeeman splittings, with fields in the range from around 1~MG to
733~MG, which amounts to 4\% of all DAs observed. As the SDSS spectra
has low signal-to-noise ratios, we carefully investigated
by simulations with theoretical spectra how reliable our detection of magnetic field was.
\end{abstract}


\section{Introduction}
In the latest white dwarf catalog based on the Sloan Digital Sky
Survey (SDSS) Data Release~7 (DR7), \citet{dr7} classify the spectra
of 19\,713 white dwarf stars, including 12\,831
hydrogen atmosphere white dwarf stars (DAs) and 922 helium
atmosphere white dwarf stars (DBs). 
The authors fit the
optical spectra from 3\,900\AA\ to 6\,800\AA\ to DA and DB grids of
synthetic non-magnetic spectra derived from model atmospheres
\citep{Koester10}.  
The SDSS
spectra have a mean g-band signal--to--noise ratio S/N(g)$\approx 13$
for all DAs, and S/N(g)$\approx 21$ for those brighter than g=19.

Through visual inspection of all these spectra, we identified Zeeman splittings
in the spectra of 521 DA white dwarfs, eleven with multiple spectra. The main 
object of this paper is to identify these stars and estimate their
magnetic field.
Independently, \citet{Kulebi} found 44 new magnetic white dwarfs in
the same SDSS DR7 sample, and used $\log g=8.0$ models to estimate the
fields of the 141 then known magnetic white dwarfs, finding fields from B=1~MG
to 733~MG. We report here on the estimate of the Zeeman splittings in
$\simeq 4$\% of all DA white dwarf stars. 
With the low resolution ($R\simeq 2\,000$) of
the SDSS spectra, magnetic fields weaker than 2~MG are only
detectable for the highest S/N spectra \citep[e.g.][]{Tout2008}. 
We first summarize
some previous results on magnetic white dwarfs.

\section{Magnetic white dwarfs and their progenitors}

The magnetic nature of the until then unexplained spectra of the white
dwarf GRW+70.8247 was confirmed by \citet{Kemp1970}.  His
magneto-emission model, which predicted the level of continuum
polarization, was not quite adequate for the high magnetic field in
this star, and the estimated field strength of 10\,MG later turned out
to be much too low, but the general idea that the strange spectrum was
caused by a magnetic field was correct. The detailed description of
the spectra became possible only  with extensive calculations of the
atomic transitions of hydrogen developed by \citet{Roesner1984} and
\citet{Greenstein1985}, and a consistent atmospheric modeling by
\citet{Wickramasinghe1988} and \citet{Jordan1992}. 

The numbers of known magnetic white dwarfs has increased significantly
since the first identifications.  \citet{Liebert2003} found that only
2\% of the 341 DAs and 15 DBs in the Palomar-Green Survey (PG) were
magnetic, i.e., exhibited Zeeman splitted lines. However, they
estimated that up to 10\% could be magnetic, if the magnetic white
dwarfs are more massive than average white dwarfs and therefore
had smaller radius and luminosities, as indicated by
\citet{Liebert88} and \citet{Sion88}.  \citet{Kawka} estimated up to
16\% of all white dwarfs may be magnetic. \citet{Jordan}, based on
spectropolarimetry using the 8.2~m telescope VLT at ESO estimate up to
15 to 20\% of all white dwarfs are magnetic at the kG level.
\citet{Landstreet12} reanalyzed the spectropolarimetry with a new state-of-the-art
calibration pipeline and added further new observations. From the total sample of 
35 DA stars they found that about 10\%\ (between 2.8\% and 30\%\ at the 95\%\ confidence level)
were magnetic on the kG level.

An accurate estimate of this percentage is crucial for an
understanding of the origin of the magnetic fields.
In the local 20 and 25~pc volume limited sample, there are
$\simeq 7\%$ magnetic white dwarfs, according to
\citet{20pc} and \citet{25pc}.

Historically, the explanation of the magnetic fields in white dwarfs
has been as fossil fields, motivated by the slow Ohmic decay in
degenerate matter \citep[e.g.][]{Braithwaite, Tout, Wickra2005}. From
the discovery of kilo Gauss (kG) magnetic fields in the atmospheres of
peculiar early type stars, the Ap and Bp stars, already \citet{Babcock1947,
Babcock1947a} demonstrated that conservation of the magnetic flux during
the stellar evolution could lead to field strengths as high as a
million Gauss (MG) in the white dwarfs resulting from the evolution.

Ap/Bp stars constitute less than 10\% of all intermediate
mass main sequence stars \citep[e.g.][]{Power}, and can account
for a fraction of 4.3\% magnetic white dwarfs, but they should
produce white dwarfs with fields above 100~MG \citep{Kawka} if magnetic flux is 
fully conserved during stellar evolution. 
\citet{Wickra2005}, via population synthesis, conclude that the current
number distribution and masses of high-field magnetic white dwarfs
(HFMWDs, $B>1$~MG) can be explained if $\approx 40$\% of
main-sequence stars more massive than 4.5~$M_\odot$ have magnetic fields in
the range of 10-100~G, which is below the current level of detection.

\citet{Schmidt1995, Liebert88, Sion88, Liebert2003}
and \citet{Kawka2007} find that magnetic white dwarfs are
more massive than non-magnetic ones by
 fitting the wings of the spectral lines to theoretical spectra,
supposedly unaffected by the magnetic fields.
\citet{Tout1995, Tout2008} and
\citet{2011arXiv1108.4849N} propose white dwarfs with fields above 1~MG
are produced by strong binary interactions during post-main sequence
evolution, while \citet{GarciaBerro12} proposes that high magnetic field white
dwarfs are produced by the merger of two degenerate cores and that the
expected number agrees with observations.
These proposals are in line with the cited observation that magnetic white
dwarf stars have, in general, higher masses then average single white dwarf stars.
\citet{Kundu12} and \citet{Das12} propose highly magnetic white dwarfs 
could have limiting masses substantially higher than the Chandrasekhar limit.
On the other hand, \citet{Wegg12} conclude that the kinematics of massive white dwarfs are consistent with the majority
being formed from single star evolution.

\section{Detection of Magnetic Field in SDSS DR7 WDs}

\begin{figure}
\includegraphics[width=\textwidth]%
{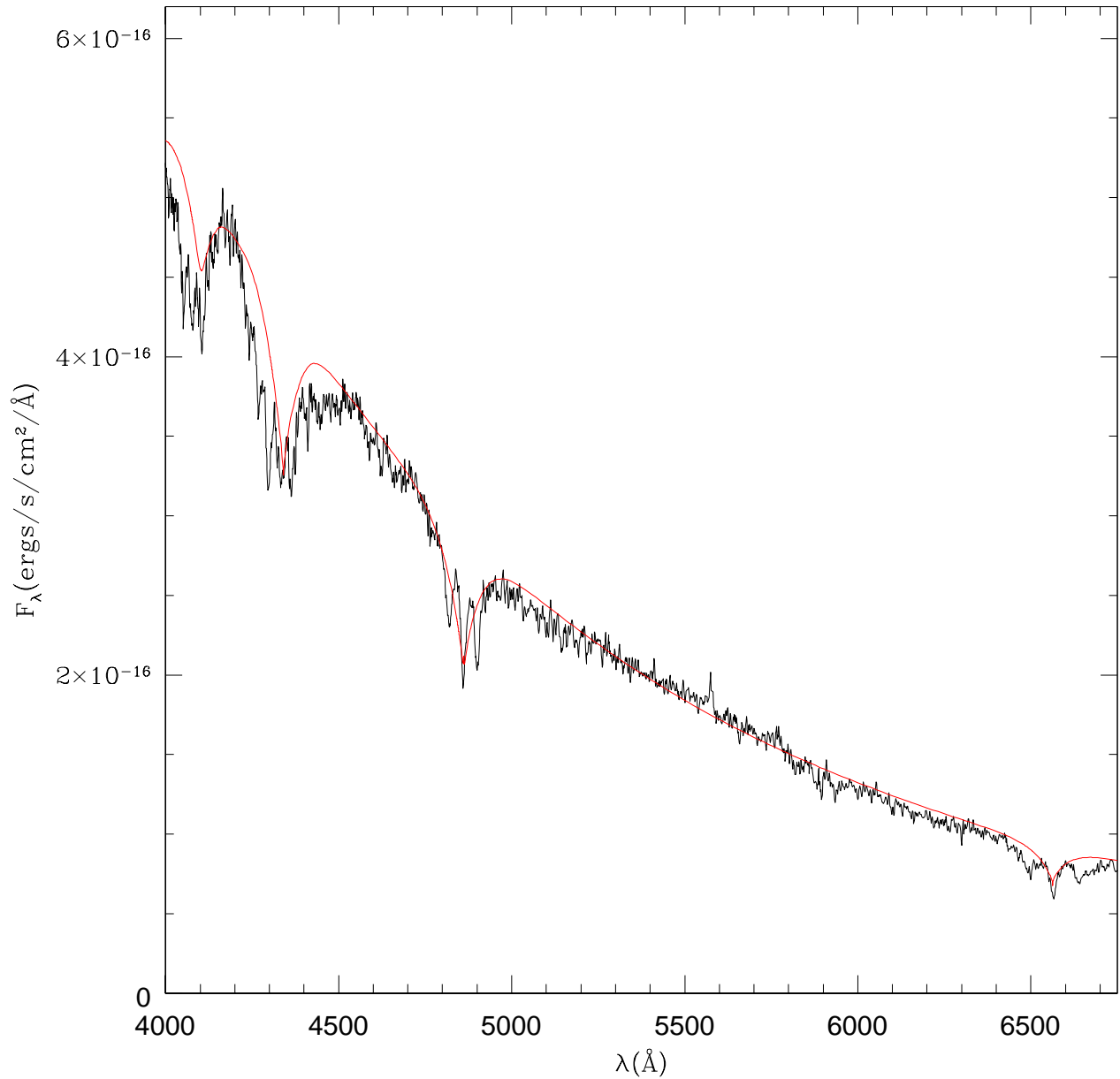}
\caption{SDSS spectrum of one of the stars we identified 
Zeeman splittings, SDSS J111010.50+600141.44, indicative of a 6.2~MG magnetic field.
A DA model {\it without magnetic field}
of $T_\mathrm{eff}=36\,000$~K, $\log g=9.64$, M=1.33~$M_\odot$ results from
a least-squares fit to the spectra, plotted in red, obviously inadequate;
the lines are wide because of the Zeeman splittings, not due to large
pressure (gravity) broadening.
\label{sdssm}}
\end{figure}

\begin{figure}
\includegraphics[width=15cm]%
{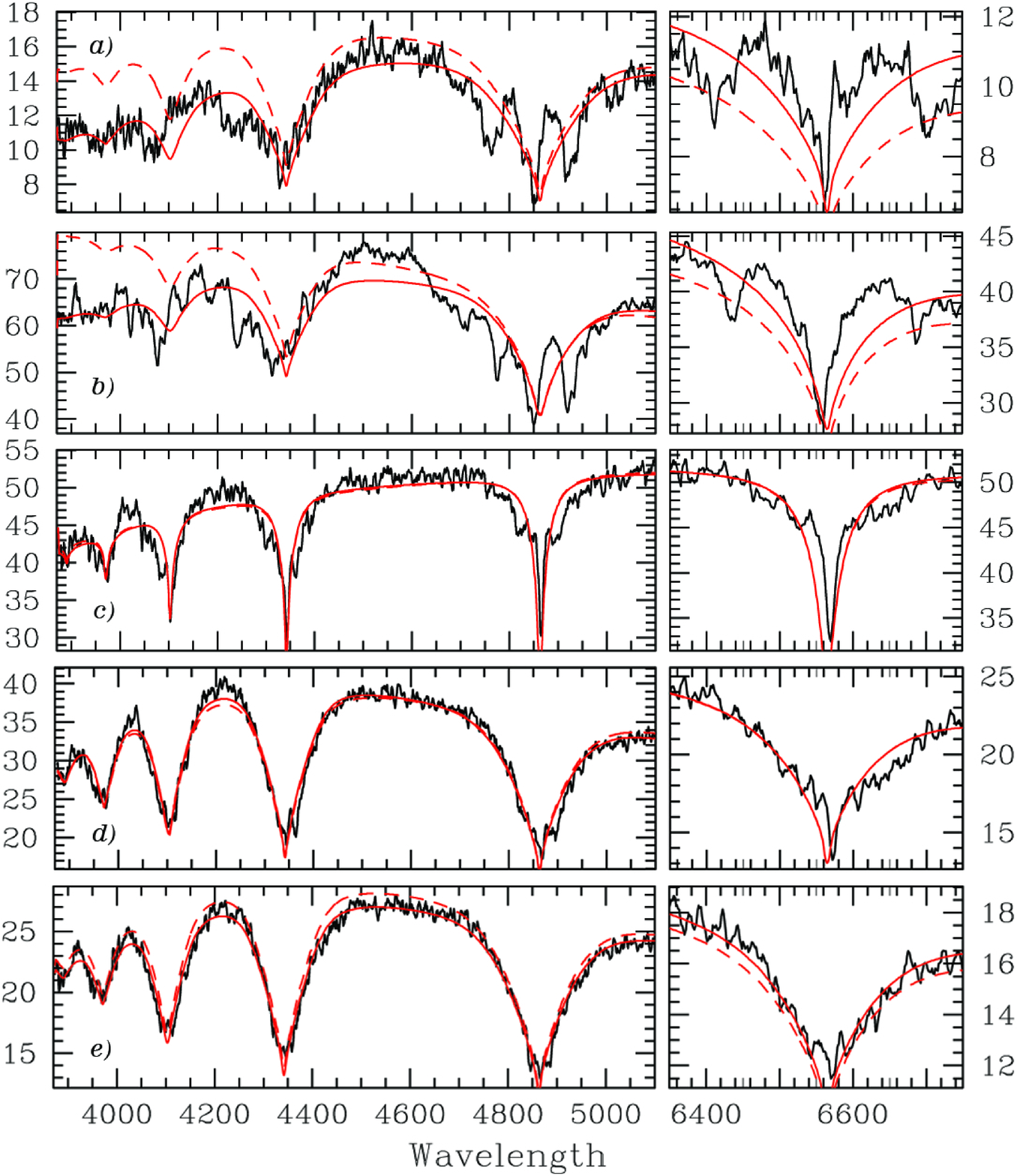}
\label{mag1}
\end{figure}

\begin{figure}
\includegraphics[width=15cm]%
{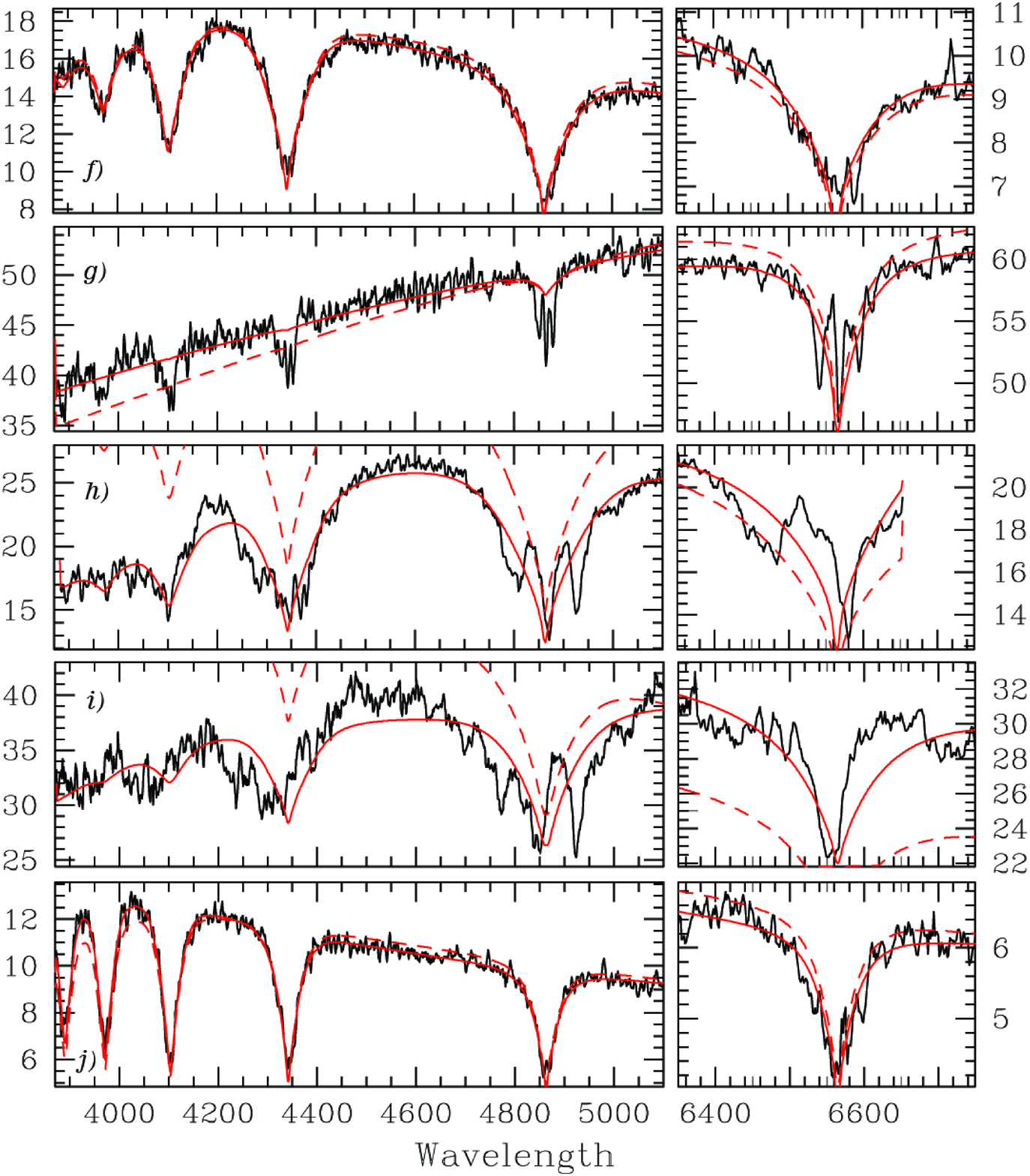}
\label{mag2}
\end{figure}

\begin{figure}
\includegraphics[width=15cm]%
{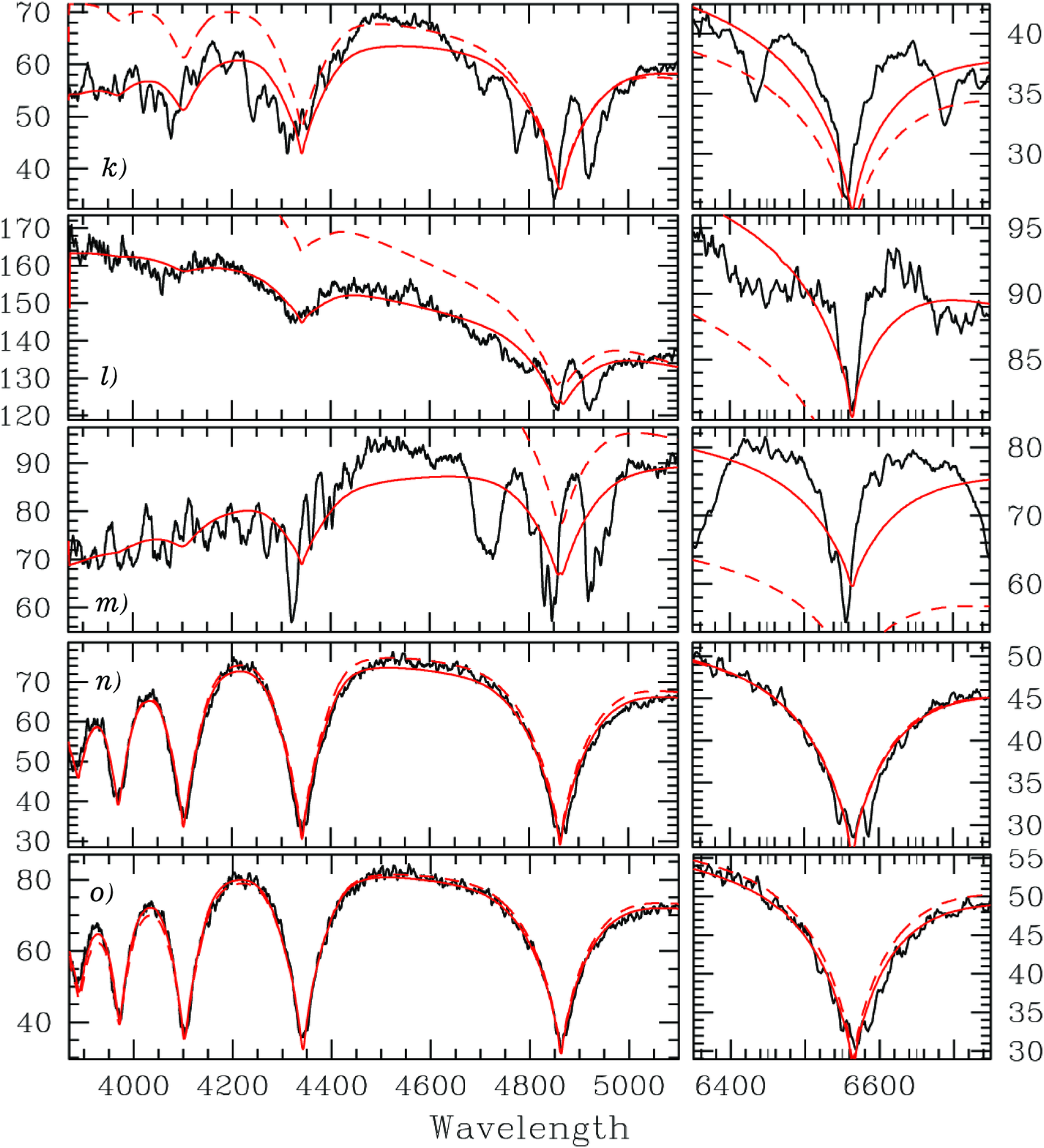}
\caption{Highest S/N SDSS spectra of a sample of stars for which we identified 
Zeeman splittings indicative of fields from 14~MG to 1.3~MG magnetic fields.
a) Plate-MJD-Fiber=1954-53357-393, 13  MG, SDSS~J101428.10+365724.40,
b) 0415-51879-378, 11~MG, J033145.69+004517.04,
c) 1616-53169-423, 2.4 MG, 123414.11+124829.58,
d) 2277-53705-484, 2.2 MG, 083945.56+200015.76.
e) 2772-54529-217, 2.2 MG, 141309.30+191832.01,
f) 2694-54199-175, 1.7 MG, 064607.86+280510.14,
g) 2376-53770-534, 2.6 MG, 103532.53+212603.56,
h) 2417-53766-568, 9.9 MG, 031824.20+422651.00,
i) 2585-54097-030, 14  MG, 100759.81+162349.64,
j) 2694-54199-528, 1.3 MG, 065133.34+284423.44,
k) 0810-52672-391, 11 MG, 033145.69+004517.04,
l) 1798-53851-233, 14 MG, 131508.97+093713.87,
m) 2006-53476-332, 19 MG, 125715.54+341439.38,
n) 2644-54210-167, 1.9 MG,121033.24+221402.64,
o) 2430-53815-229, 2.5 MG, 085106.13+120157.84.
DA models {\it without magnetic field} results from
least-squares fits to the spectra are plotted in red, obviously inadequate.
\label{magnetic}}
\end{figure}
\begin{figure}
\includegraphics[width=\textwidth]%
{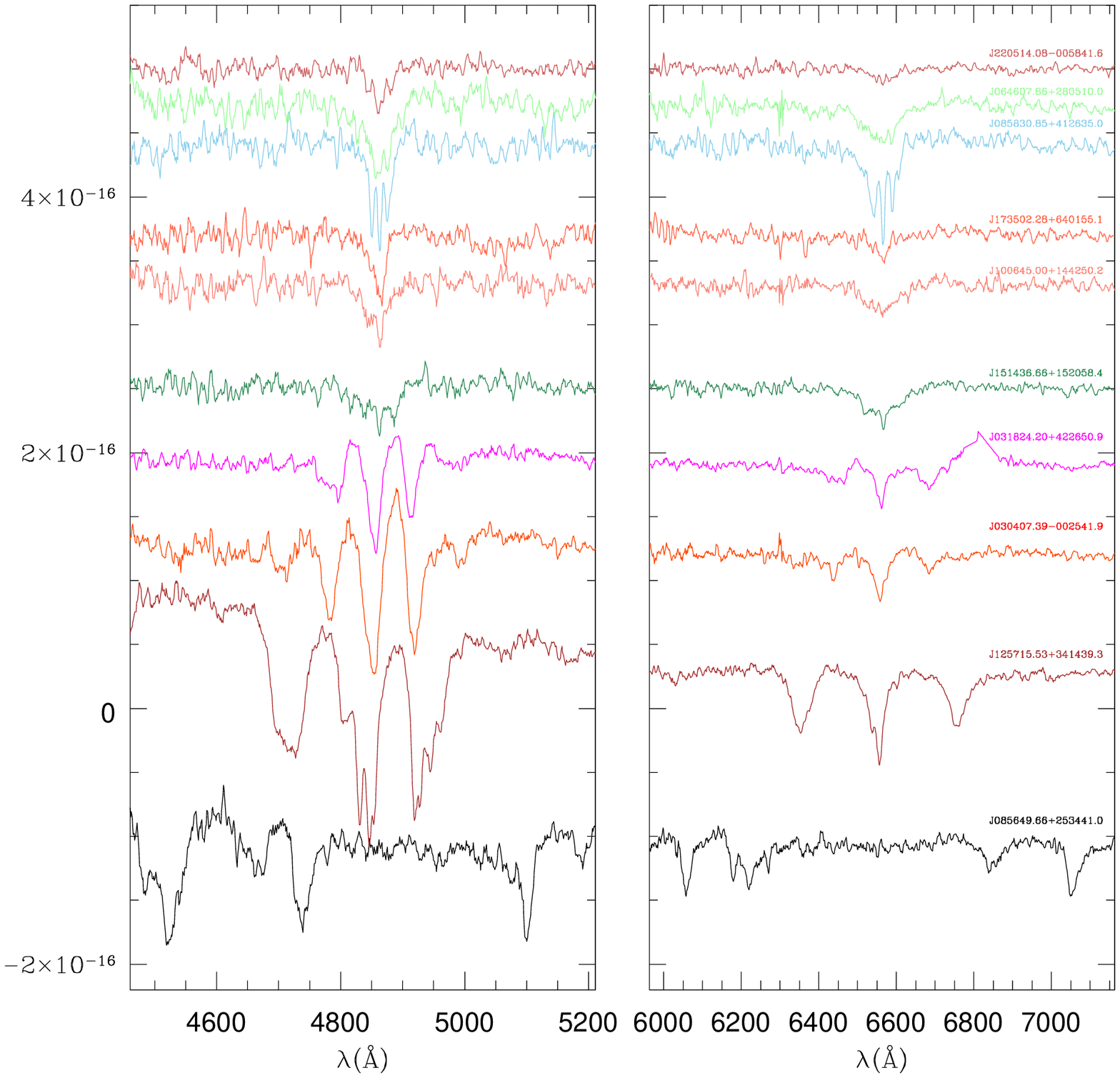}
\caption{H$\beta$ (left-hand panel) and H$\alpha$ (right-hand panel) line profiles for a sample of new magnetic white dwarf stars, 
with fields of $\simeq 3$~MG at the top, 
and 90~MG at the bottom. The y-column shows flux in arbitrary units.
\label{magc}}
\end{figure}

We classified more than 48\,000 spectra, selected as possible white dwarf stars
from the Sloan Digital Sky Survey Data Release~7 by their colors,  through
visual inspection 
and detected Zeeman splittings in 521 DA stars. 
Figure~\ref{sdssm} shows the spectra of one of the newly identified  magnetic white dwarfs as an example.
As we were only able to detect magnetic
fields down to 1--3~MG in strength,
because of the R$\simeq$2\,000 resolution and relatively low signal-to-noise of
most spectra ($\langle S/N \rangle \simeq 13$), the 4\% detected (521/12831 DAs) is a lower limit and the actual number
of magnetic white dwarf stars should be  larger if we include smaller field strengths.
The identified magnetic white dwarf stars cover the whole range of temperature and 
spectral classes observed \citep{dr7}. 
Figures~\ref{magnetic} and \ref{magc} show spectra of a few of the highest S/N new magnetic white dwarfs we identified, showing
a broad range of splittings, and hence of magnetic fields.

\begin{figure}
\centering
\includegraphics[width=0.9\textwidth]%
{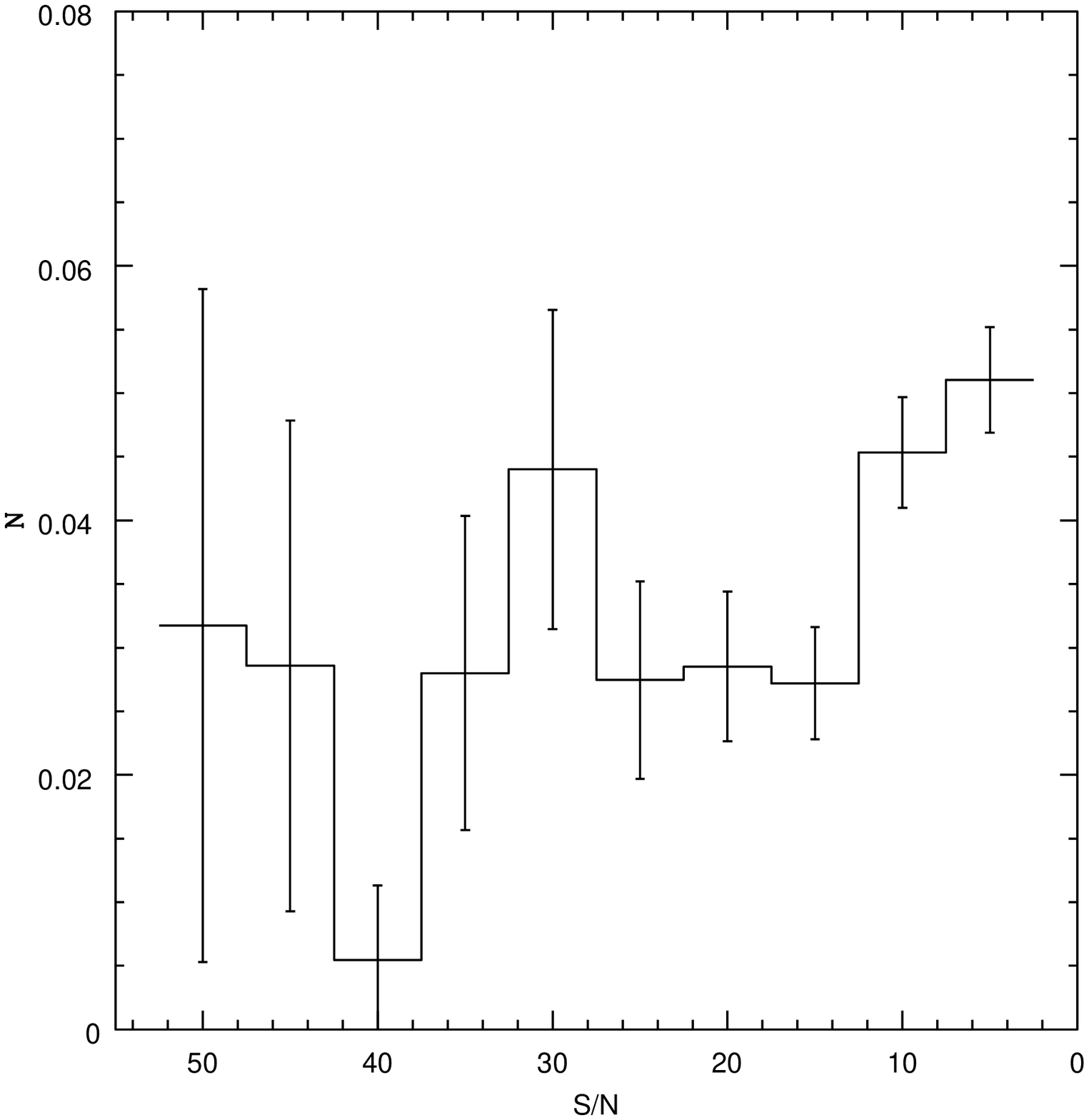}
\caption{Fraction of detected magnetic DA white dwarfs versus S/N of the spectra.}
\label{sg}
\end{figure}

Fig.~\ref{sg} shows the fraction of detected magnetism in white dwarfs  as a function of the signal-to-noise (S/N) ratio provided by \citet{dr7}.
The fact that we see an increase of detected magnetic fields in spectra with
lower  S/N ratios made us suspicious. For this reason we carefully investigated the influence of the S/N ratio on the
detection rate with the help of a blind test using noisy theoretical spectra (see Sect.\,\ref{blind}).
Our result was that classification with S/N $\leq$10 need to be confirmed by
future observations.
 Furthermore, any estimate of the overall percentage of magnetic to
 non-magnetic white dwarf stars needs to take this apparent selection
 effect into account  \citep{Liebert2003}.

\subsection{Estimation of  the Magnetic Field Strength}

\begin{figure}
\includegraphics[width=\textwidth]%
{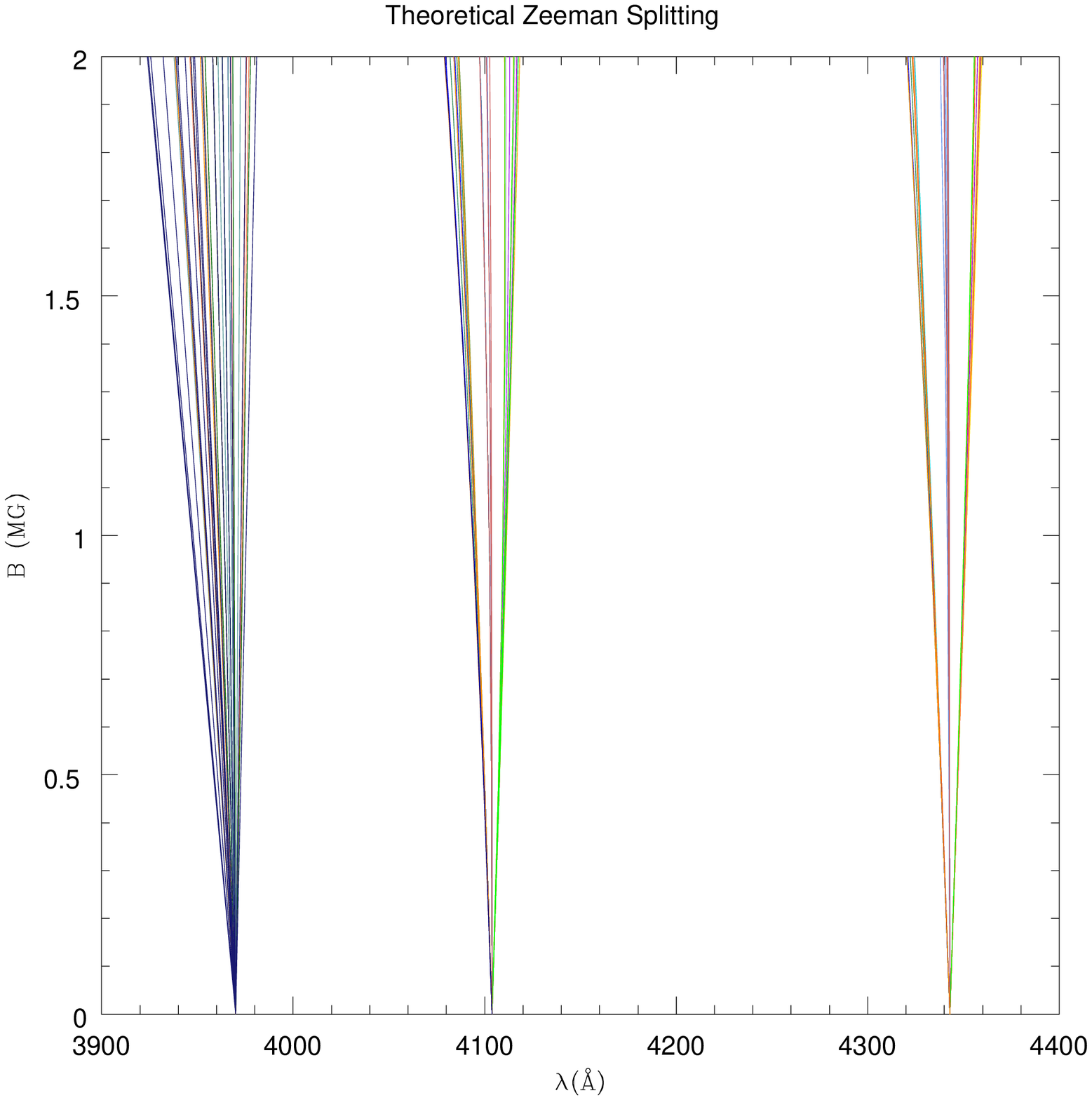}
\caption{H$\epsilon$, n=$7\rightarrow 2$, $\lambda_0=3971$~\AA\ (left) 
to H$\gamma$, n=$5\rightarrow 2$, $\lambda_0=4342$~\AA\ (right) theoretical
Zeemam splittings for B=$0\rightarrow 2$~MG, showing the higher lines
split into multiplets even for these low fields, because the quadratic Zeeman splitting
is proportional to $n^4$.
\label{kulebih}}
\end{figure}

For fields stronger than 10~kG but weaker than 2~MG, i.e.,
in the Paschen-Back limit,
low level ($n\leq 4$) lines will be split into three components, with the shifted components
separated by around 
\[\Delta \lambda = \pm 4.67\times 10^{-7} \lambda^2 B\]
with $\lambda$ in {\AA}ngstrons and B in MG
\citep{Jenkins1939,Hamada1971,Garstang},
The quadratic splitting is given by
\[\Delta \lambda_q = -\frac{e^2}{a_0^2}{8mc^3 h}\lambda^2 n^4 (1+m_\ell^2)B^2
\simeq -4.97 \times 10^{-23} \lambda^2 n^4 (1+m_\ell^2)B^2~\mbox{\AA}\]
where $a_0$ is the Bohr radius, and $m_\ell$ the magnetic quantum number.
This formula is valid for $2p$ to $ns$ and $2p$ to $nd$ transitions,
where $n$ is the principal quantum number. 
For the $2s$ to $np$ transitions,
\[\Delta \lambda_q \simeq -4.97 \times 10^{-23} \lambda^2 [n^2(n^2-1)(1+m_\ell^2-28)B^2~\mbox{\AA}\]
Note that because of the $n^4$ dependency of the quadratic Zeeman splitting,
even for fields around 1~MG, the $n\geq 7$ lines show dominant
quadratic splittings (Fig.~\ref{kulebih}).

For magnetic fields less than $\simeq$~2~MG, the Zeeman splitting is difficult
to observe in low resolution spectra of
white dwarfs because the spectral lines are already
broadened due to the high density.
The linear Zeeman splitting is equivalent to a broadening of unpolarized
spectral lines of the order of 10~km/s for fields around 10~kG.
For higher fields the magnetic energy cannot be included as a perturbation
because the cylindrical symmetry of the magnetic field start to
disturb the spherical symmetry of the Coulomb force that keeps
the hydrogen atom together. For the $n=1$ level, the Lorentz force and the
Coulomb force are of the same order for B=4\,670~MG.
As the energy of the levels is proportional to the inverse of $n^2$,
the higher levels are disturbed for much smaller fields.

The observed Zeeman splitting represents the mean field across the
surface of the star. If the field is assumed as dipole, the
mean field is related to the polar field by
\[B = \frac{1}{2} B_p \sqrt{1+3 \cos^2 \theta}\]
where $B_p$ is the polar field and $\theta$ is the angle
between the field and the line of sight. Simple centered dipoles
are rarely, if ever, seen in real stars \citep[e.g.][]{Kulebi}.

To estimate the magnetic fields, we measured the H$\alpha$ and H$\beta$ mean splittings
independently and used the mean fields estimated by \citet{Kulebi} as scale. 
Our measurements are of the mean line centers, by visual inspection,
and therefore do not take into account the shape of the lines,
which are different due to the fact the for most stars the
magnetic field is not centered at the center of the star \citep{Kulebi}.
Our estimates also ignore any effects due to higher moments than dipoles, or double-degenerate stars.
The estimates are therefore very rough, but do indicate the order of magnitude of the
magnetic field.

\begin{figure}
\centering
\includegraphics[width=0.9\textwidth]%
{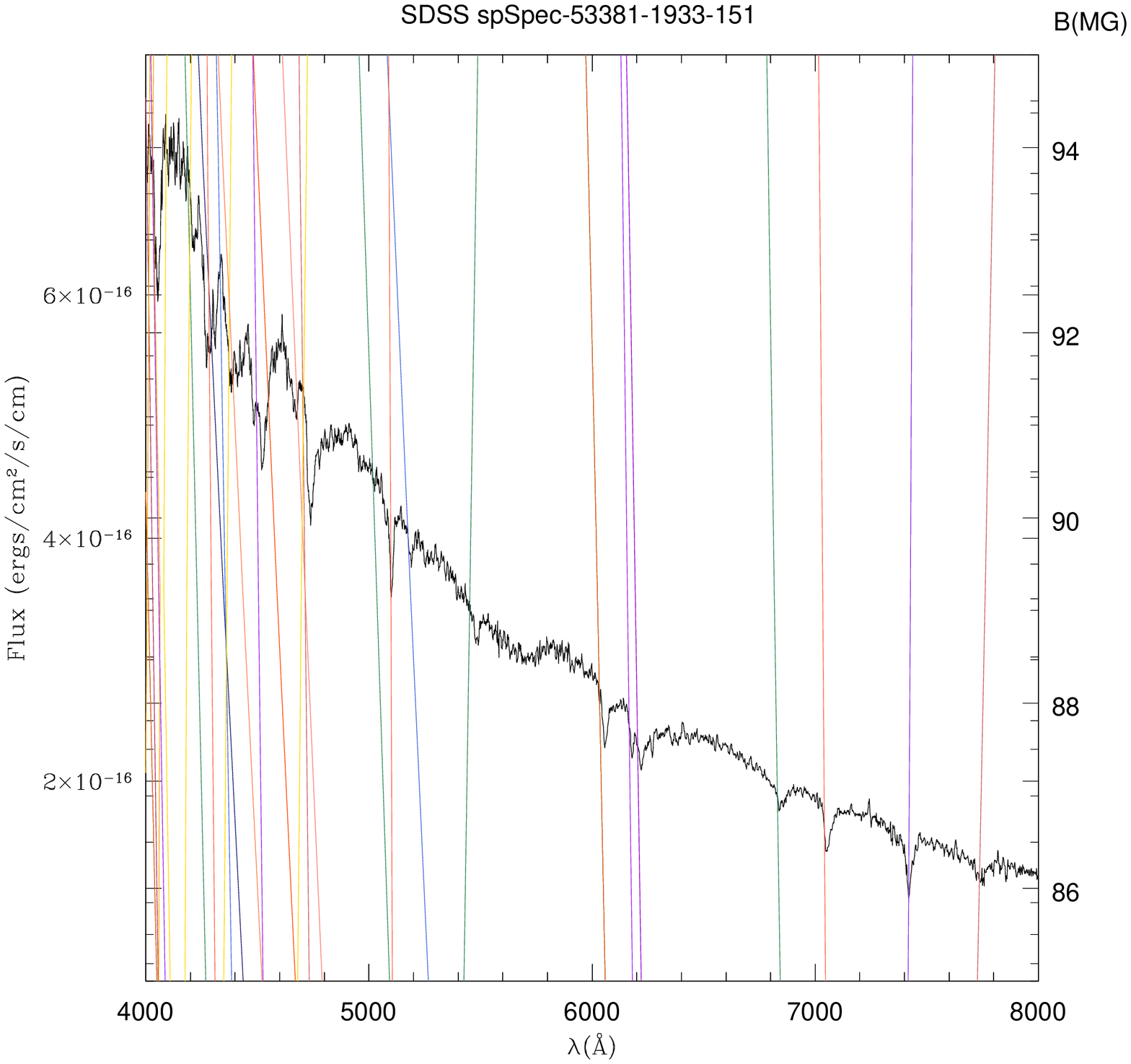}
\caption{Spectrum for SDSS~J085649.68+253441.07 DA with a magnetic field around 90~MG,
and the position of the theoretical Zeeman splittings (continuous colored lines) for a dipole magnetic field B indicated on the right of the plot.}
\label{b90}
\end{figure}

For fields above 30~MG, like for SDSS~J085649.68+253441.07 shown in Figure~\ref{b90},
the line identification is difficult, and we adjusted graphically the spectra to the
theoretical Zeeman positions only.

\begin{figure}
\centering
\includegraphics[width=0.45\textwidth]{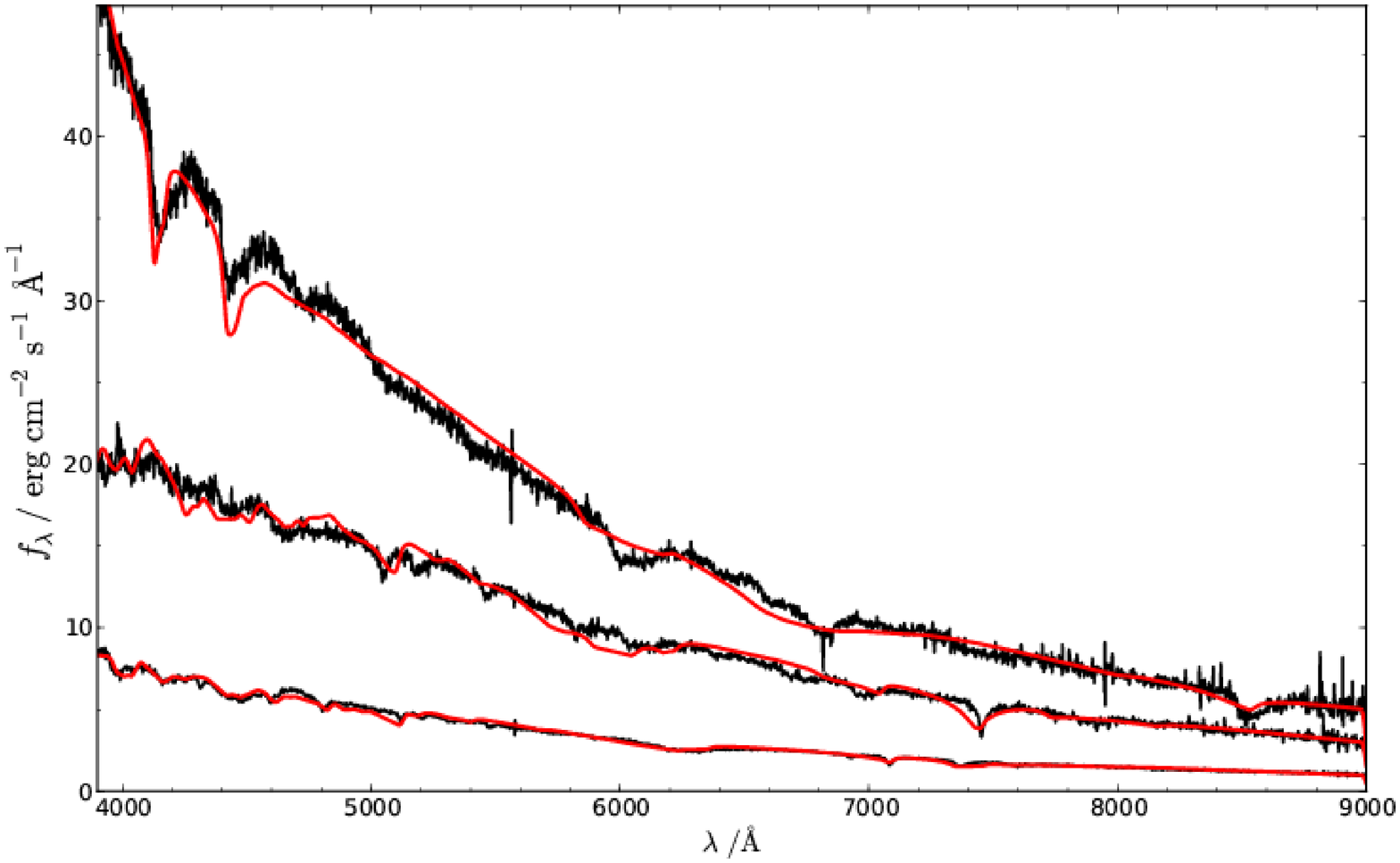}
\includegraphics[width=0.45\textwidth]{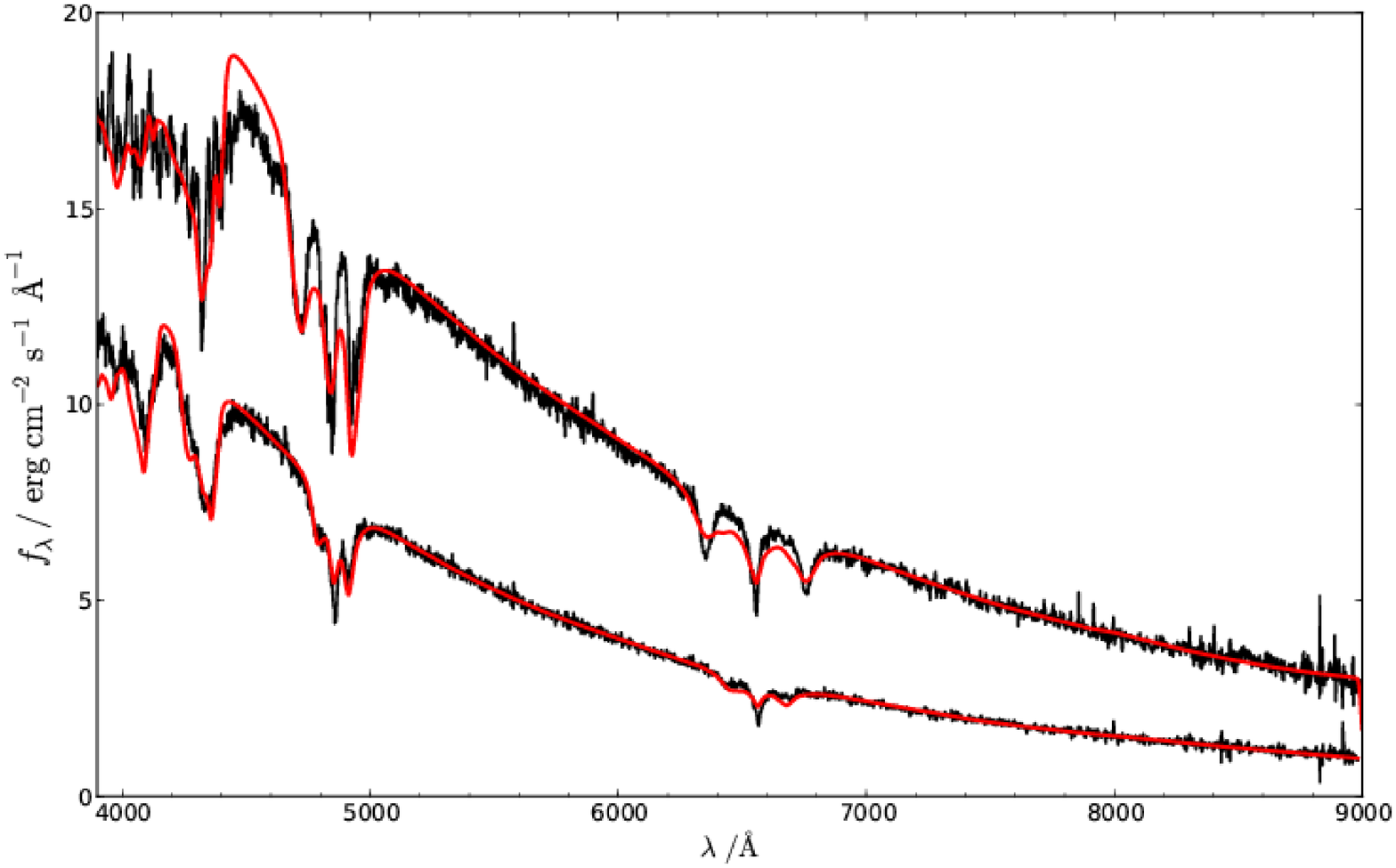}
\caption{Sample fits done to 5 different objects using the method from \citet{Kulebi},
 with the database of centered dipolar models. The  plots
 include the following observed spectra (black lines) with the best fit dipole models (in
 red lines): (first part) SDSS J135141.13+541947.35 with ($B_p$ = 500 MG), SDSS
 J021148.22+211548.19 ($B_p$ = 168 MG), SDSS J101805.04+011123.52 ($B_p$ = 127
 MG);  (second part) SDSS J125715.54+341439.38 ($B_p$ = 12 MG), SDSS
 J074853.08+302543.56 ($B_p$ = 6.8 MG). The fits are intended to be
 representative and the disagreements between model and fits are due to a lack
 of detailed modeling in which effective temperature and the sophisticated
 magnetic models has not been accounted for.
In the plots, arbitrary factors of
 normalization have been used for display.
\label{baybars}
}
\end{figure}

Fig.~\ref{baybars} shows fits of centered dipole magnetic models with
$\log g=8.0$ as those by \citet{Kulebi} for 5 stars, to illustrate the discrepancies of assuming centered fields.
Table~1 shows the estimated values for the magnetic fields for the 521 spectra we measured.
The 5th column of the table shows the signal-to-noise ration in the region of the $g$
filter of the spectra, S/N(g). 

\begin{figure}
\centering
\includegraphics[width=0.8\textwidth]%
{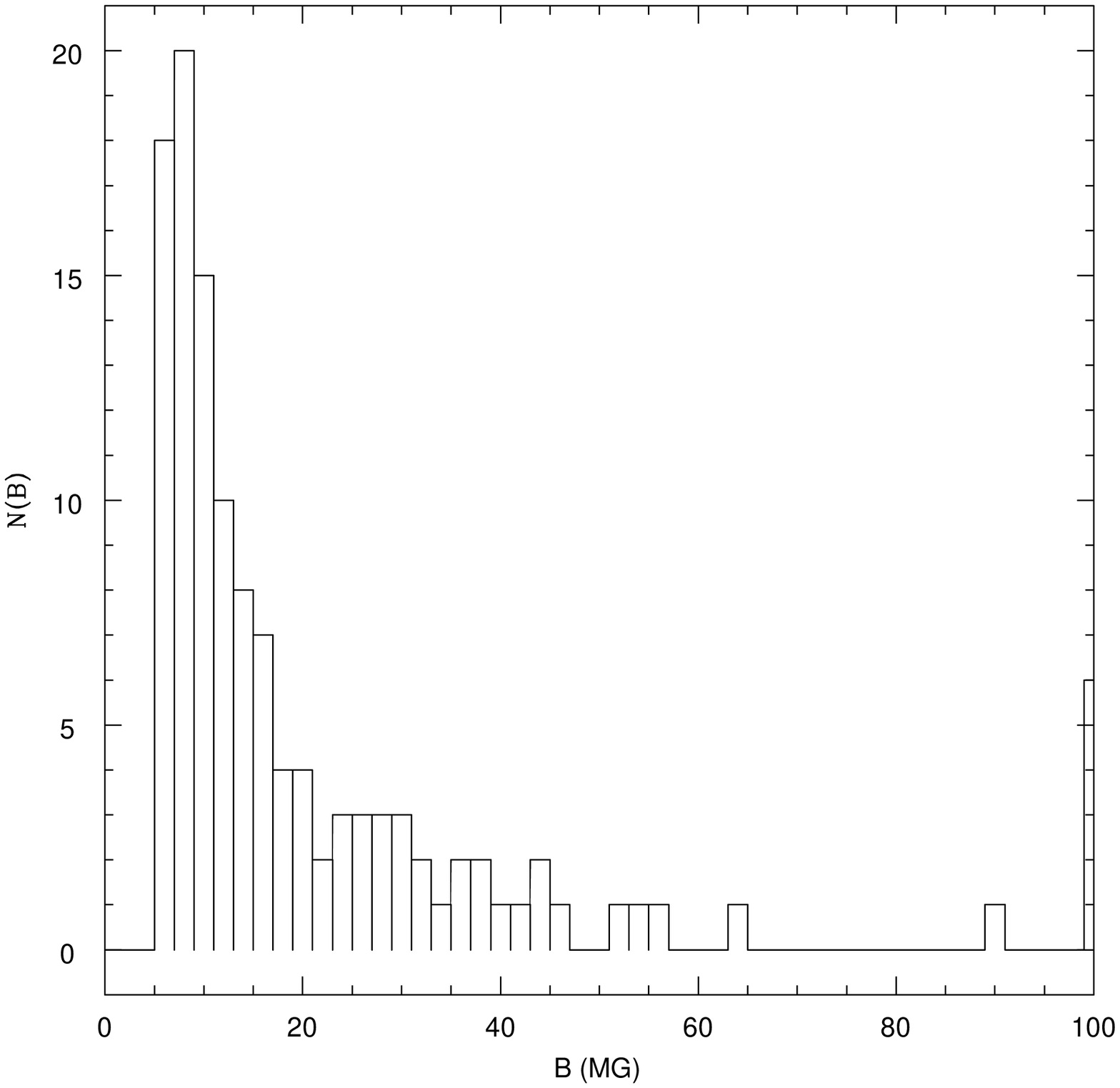}
\caption{Number of white dwarfs versus magnetic field for the SDSS sample.
As our detection limit is around 1--2~MG, selection effects in the lowest bin are important.} 
\label{field}
\end{figure}

Fig.~\ref{field} shows the distribution of fields for our sample,
showing an increase in the number of stars for lower fields, 
except for the lowest bin, where selection effects are important,
as our $R\simeq 2000$  resolution implies we cannot detect $B\leq 2$
except at the highest S/N.

\section{Blind test}\label{blind}
In order to check whether magnetic white dwarfs can be identified and analyzed
with sufficient confidence using noisy spectra, we have performed a blind test.
One group has
calculated model spectra for white dwarfs with and without magnetic fields for
effective temperatures between 8000 and 40000\,K and  $\log g=8.0$.
For the models with magnetic fields we assumed centered magnetic dipoles with a
polar field strength between 1 and 550\,MG and viewing angles between the
observer and the magnetic fields between 0 and $90^\circ$.  Subsequently, we
added Gaussian noise with signal-to-noise ratios between 4 and 35.

In total 346 such spectra were given to the  second group whose task it was to
identify which of the objects were magnetic and what the mean magnetic field
strength was. This group did not know how many of the spectra were calculated
assuming no magnetic fields and what the assumed field strength for the magnetic
objects were.

To be on the secure side, we assumed that all objects with a magnetic
field lower than 2\,MG were regarded as non magnetic.

79 of the 346 noisy spectra were based on  zero-field models ($<2\,$MG).  Only
seven of them were  wrongly classified as being magnetic and all of them had
signal-to-noise ratios below eight; in total we have have simulated 43 objects with S/N$<8$.
None of the false detections had a determined field strength
above 3\,MG so that no non-magnetic white dwarf was regarded as having a
strong magnetic field. 
If we disregard detections below 2\,MG and signal-to-noise ratios below 10 we do not detect any false-magnetics.

41 of the noisy theoretical spectra were assigned to be non-magnetic by the
second group but in fact had assumed magnetic fields larger than 2\,MG. 
This number is indeed significantly large  because we had in total 130 objects with
simulated zero fields.  At $B<50$\,MG 13 out of 265 objects (5\%) were false-negatives.
At $B>50$\,MG  we have 28 out of 81 objects (34\%\ false) false negatives.  The
distribution between 50 and 400\,MG is quite flat. 
If we limit ourselves to signal-to-noise ratios above ten the
number of false-negatives is reduced to seven objects (out of 84) which all had relatively weak features
(magnetic fields above 100\,MG and effective temperature above 35000\,K). At S/N$>15$ this number is further reduced
to two (out of 52) objects; at S/N$>20$ (34 simulated objects)  all simulated magnetic objects were determined as such. 

The determined mean field strengths  were compared to the mean magnetic fields
of the dipole models.
214 of the theoretical spectra were calculated for  field strength between 1
and 100\,MG. If we  again limit ourself to the ones with signal-to-noise
ratios above ten, the magnetic field determination ``by eye'' was rather
accurate. In only six cases the  determined magnetic field strength differed
from the simulated one by more than a factor of two.

At field above 100\,MG  (53 simulated spectra) the magnetic-field determination
was less satisfactory even at signal-to-noise ratios above 15.
The magnetic fields were often wrong (mostly underestimated) by more than a
factor of two.

Without detailed modeling, the magnetic field determination at very high
magnetic fields ($>100$\,MG) is much more difficult than at lower fields. This is
because most of the spectral lines are completely washed out by the quadratic
Zeeman effect if the magnetic field varies over the stellar surface (this
variation amounts to a a factor of two in the case of dipole models). Only
the so-called
stationary line components for which the  wavelengths go through maxima or
minima as functions of
the magnetic field strength remain visible. The corresponding field strength is
not necessarily close to the mean field strength. This could partially
explain the difference between the field determinations  ``by eye'' and the
simulated values.

We conclude that we can distinguish between spectra from magnetic ($>2\,$MG)
and non-magnetic white dwarfs ($\le 2$\,MG)  with very high confidence if we
limit
ourselves to spectra with signal-to-noise ratios above 10. Hot magnetic white
dwarfs with effective  temperatures above  35000\,K and fields above 100\,MG can
be missed due to their shallow features. For field strength above 100\,MG we
generally have to assume large uncertainties in the ``by eye'' field determination.

\section{Variable Fields}
\begin{figure}
\centering
\includegraphics[width=0.9\textwidth]%
{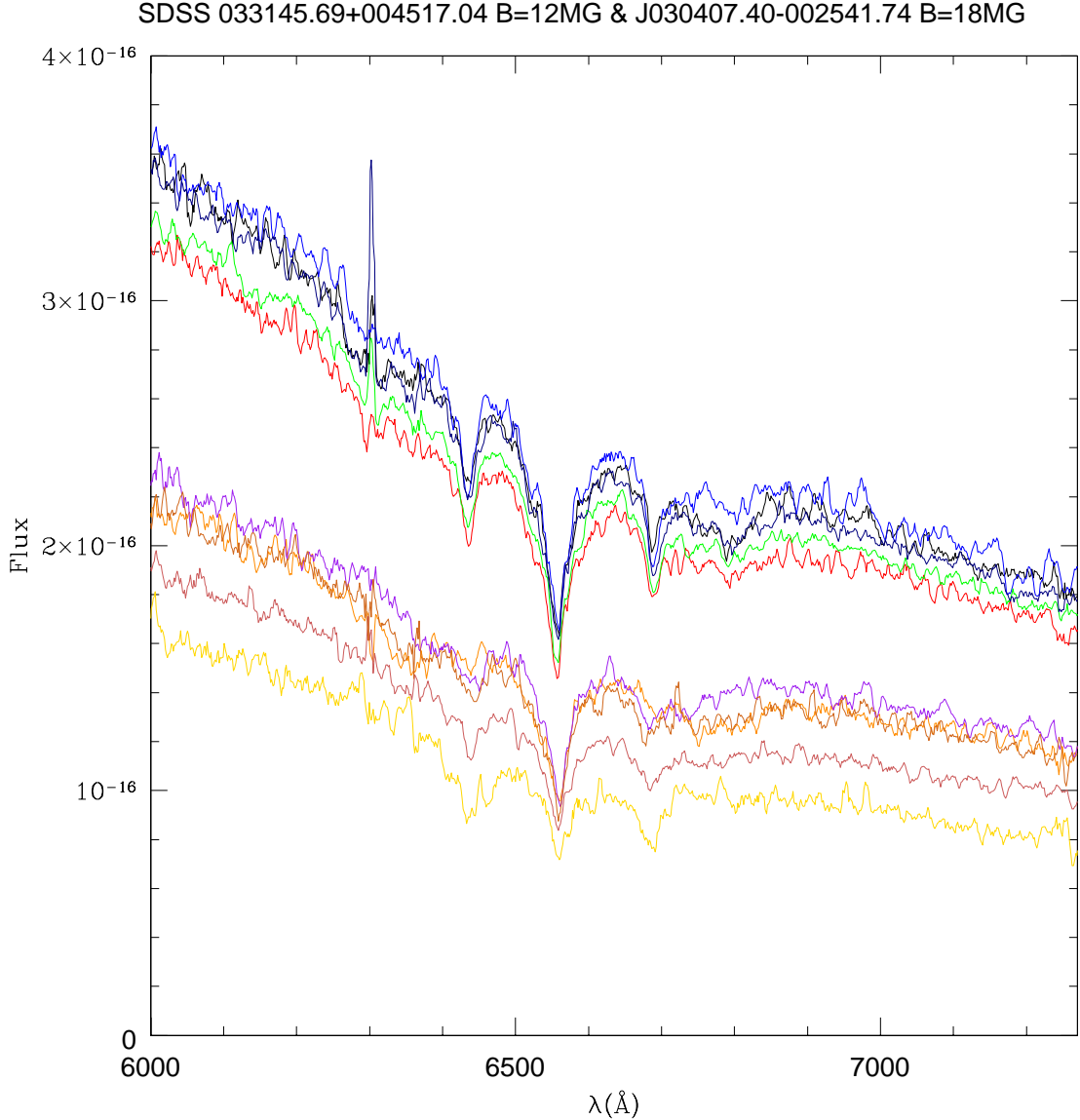}
\caption{Spectra for different epochs for SDSS~J033145.69+004517, with B=12~MG (top), and
SDSS~J030407.40-002541.74, with B=18~MG. The last star set of spectra shows significant changes in the
Zeeman splittings with time. The spectra are not purposefully offset from each other.
\label{duplicate}
}
\end{figure}

For eleven stars we have from two to six independent co-added spectra, obtained at
different epochs, and for a few of them we could see significant
changes in the Zeeman splittings, probably due to an inclined magnetic field axis with
respect to the rotational axis of the star. For SDSS~030407-002541.74,
for example, shown in Fig.~\ref{duplicate},
the structure of the Zeeman splitting changes substantially,
indicating either a very complex magnetic structure, or possibly
a double degenerate magnetic system.
We do not have time series spectra to study their variability timescale,
but such changes in the line profiles have been detected for other magnetic white
dwarfs, due to rotation \citep[e.g.][]{Burleigh99,Euchner02,Euchner05,Euchner06}.
\citet[e.g.][]{Breedt12} shows some of the magnetic white dwarfs are in fact
white dwarf binaries, when phase resolved spectra are obtained.

\section{The effect of magnetic field on mass estimates}
\begin{figure}
\centering
\includegraphics[width=\textwidth]%
{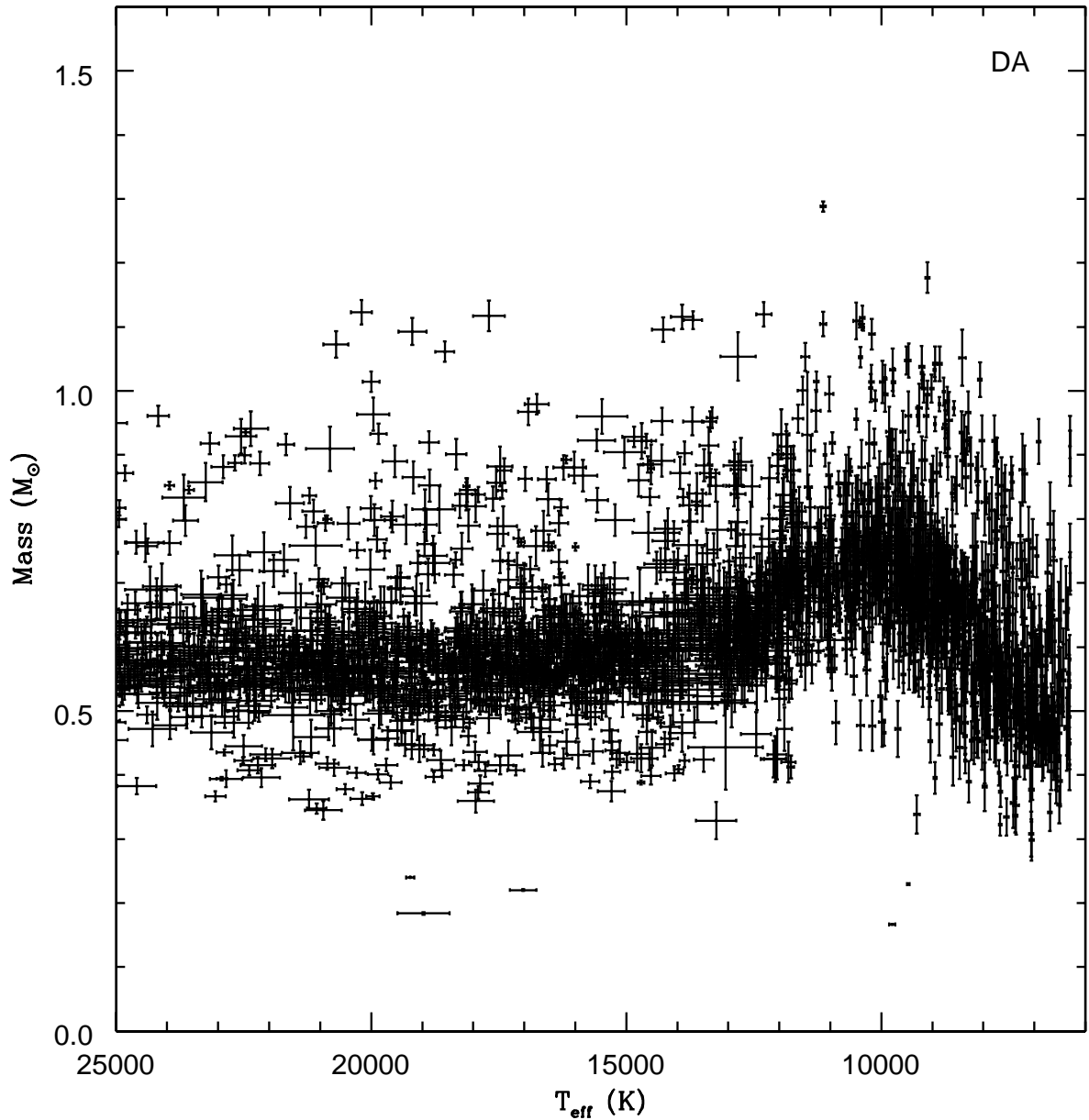}
\caption{Masses for DA stars measured from the $S/N\geq 15$ SDSS optical spectra by
\citet{dr7}, showing the apparent increase in the derived masses around
$T_\mathrm{eff}\leq 13\,000$~K, which are not seen in the masses derived
by colors or gravitational redshift.}
\label{mass}
\end{figure}

The mass distribution of the hydrogen-rich DAs 
shows an effect, which is well-documented since
many years, but still not fully understood: the average mass, as
estimated by the surface gravity, increases apparently below 13\,000K
for DAs
\citep{Bergeron91, koester91, scot,
  Liebert2005, Kepler2007, Limoges, Gianninas2010, Tremblay11, Gianninas2011}.
Single white dwarf masses in these studies are typically determined
through spectroscopy - measuring line widths due to Stark and neutral
pressure broadening. Mass determinations from photometry, and
gravitational redshift \citep{Koester06,falcon10} do not show this
mass increase, so the increase is probably not real, and merely
reflects some failure of the input physics in our spectroscopic
models.  Efforts have been made to improve the treatment of the line
broadening \citep{Koester2009, Tremblay10}, but the apparent mass
increase remains \citep{Gianninas2010, Tremblay11, Gianninas2011}.
In Fig.~\ref{mass} we show the masses for DAs with S/N$\geq$15
spectra in \citet{dr7}.

Other proposed explanations for the broadening were the treatment of the hydrogen level
occupation probability, or convection bringing up subsurface He to the
atmosphere, increasing the local pressure. However, no evidence for
the He could be found, leaving the very description of convection with
the usual mixing length approximation as the most likely culprit
\citep{Koester2009, Tremblay10}. Calculations 
using realistic 3D simulation of convection seem to confirm
this assumption \citep{Tremblay11a}.

\begin{figure}
\centering
\includegraphics[width=0.8\textwidth]{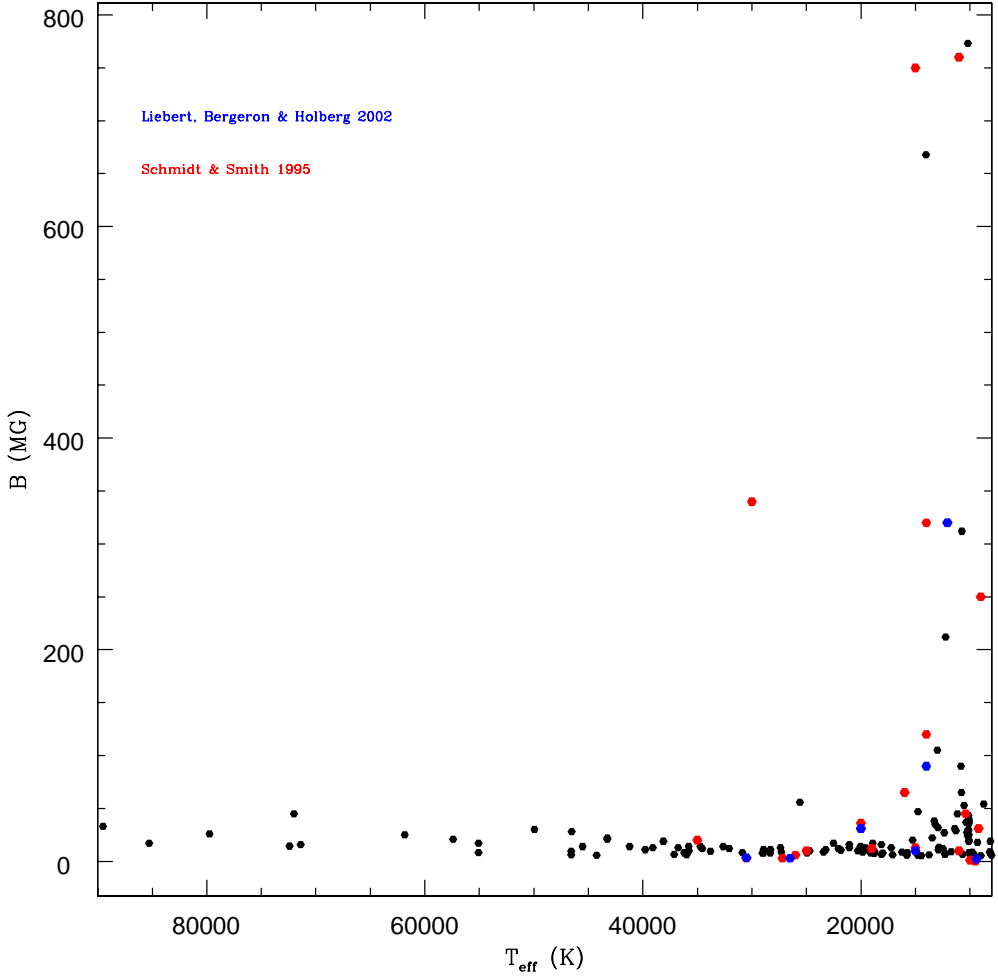}
\caption{Magnetic field versus effective temperature for the SDSS sample, 
\citet{Liebert2003}, and \citet{Schmidt1995}, showing an increase in field
for $T_\mathrm{eff}\leq 13\,000$~K, where a surface convection zone develops.
As the number of stars is larger at lower temperatures just because they cool
on a longer timescale, the fraction of higher field stars is the important
parameter.
\label{DAH}}
\end{figure}

\begin{figure}
\centering
\includegraphics[width=0.8\textwidth]{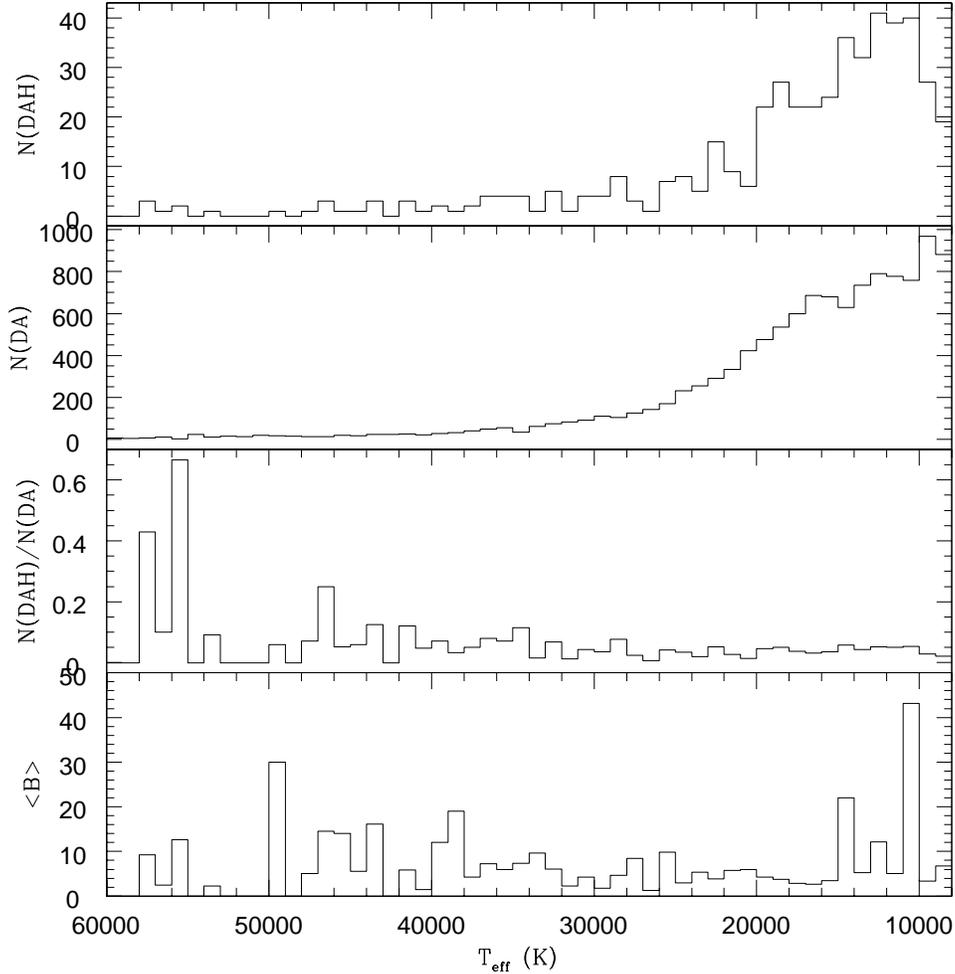}
\caption{The two top panels show the number of DAHs and DAs versus $T_\mathrm{eff}$.
The third panel from the top shows the ratio of the number of DAHs to DAs.
On the lowest panel, we see the mean field for DAHs of
that $T_\mathrm{eff}$, showing that even though the fraction of
magnetic to non-magnetic measured with $B>2$~MG did not increase at lower
$T_\mathrm{eff}$, the mean field does increase.
\label{DAH1}}
\end{figure}

In this study we explore a complementary possibility for the broadening
of the spectral lines below $T_\mathrm{eff}\simeq 13\,000$~K, the presence of weak
magnetic fields in the cooler white dwarfs. Unresolved Zeeman
splitting can increase the apparent line widths and mimic a stronger
Stark broadening, specially for the higher lines. Since the spectroscopic gravity determination is
based on the line widths, an average mass white dwarf star with a
weak magnetic field can appear spectroscopically indistinguishable
from a non-magnetic, massive star. The combined effect of electric and
magnetic fields on the spectral lines is very complicated and has been
studied only for special cases of the geometry
\citep[e.g.][]{Friedrich94, Kulebi}. Detailed model grids, which
include also the effect of the magnetic field on the radiative
transfer are therefore not yet available. 
We find an increase in the mean field around the same temperature when these
stars develop a surface convection zone, raising the possibility that the
surface convection zone is amplifying an underlying magnetic field 
(Figures \ref{DAH} and \ref{DAH1}).

As the Zeeman splitting broadens the lines, we cannot use the line
profiles to estimate their surface gravity directly. For fields stronger than
$B\simeq 1$~MG, the magnetic splittings for the n=7--10 energy levels of hydrogen washes out the lines, just like high gravity does.
As shown in Fig.~\ref{kulebih} the theoretical Zeeman splittings for
the Balmer lines initiating at
n=7 (H$\epsilon$, $\lambda_0=3971$~\AA) to n=5 (H$\gamma$), showing the higher lines split into multiplets even
for fields below $B=1$~MG.
Unfortunately there are no published calculations of the splittings for 
hydrogen levels higher than n=7 for these fields, where perturbation
theory is no longer applicable \citep{Jordan1992,Ruder1994}. 
For these higher
levels, the Zeeman splittings calculations need higher order terms even for fields of the
order of 1~MG.

As we detected Zeeman splitting in the
disk integrated spectra for 4\% or more of white dwarfs, 
which comes from global organized fields, perhaps even smaller or
unorganized fields are the cause for the line broadening
on these cooler stars.

It will be necessary to investigate if surface convection amplification of an 
underlying weak magnetic field is causing broadening of the spectral lines of 
white dwarf stars cooler than ~13\,000~K, leading to misinterpretation of
these stars as more massive stars.

Even weaker magnetic fields in white dwarfs have been studied by
\citet{Koester1998}, who
obtained high resolution
spectra measurements of the NLTE core of H$\alpha$ for
28 white dwarf stars to measure their projected 
rotational velocities, finding 3 magnetic
white dwarfs,
and no fields above 10-20~kG for the other stars,
all hotter than 14\,000~K. 
\citet{Koester2009AA}
observed about 800 white dwarfs in the SPY survey,
finding 10 magnetic, with fields from 3 to 700~kG.
\citet{Kawka12} studied 58 white dwarfs with ESO/VLT/FORS1 spectropolarimetry
and estimate 5\% of white dwarfs with fields 10 to 100~kG.
\citet{Landstreet12} estimate that 10\%\ of all
white dwarf stars have kG fields from ESO/VLT/FORS spectropolarimetry.

\section{Mean Masses}
\citet{Wickra2005} quote a mean mass of 0.93~$M_\odot$  for magnetic white dwarfs,
based on \citet{Liebert2003} determinations, 
but the sample includes 
only a handful of stars with astrometric measured masses,
so the evidence that the magnetic white dwarfs
are more massive than the average were scarce.
\citet{Kawka2007} obtained a mean mass of $0.78~M_\odot$ for the 28 magnetic DA white dwarfs with mass estimates,
mainly from fitting the line wings of the spectra.

\begin{figure}
\centering
\includegraphics[width=0.9\textwidth]%
{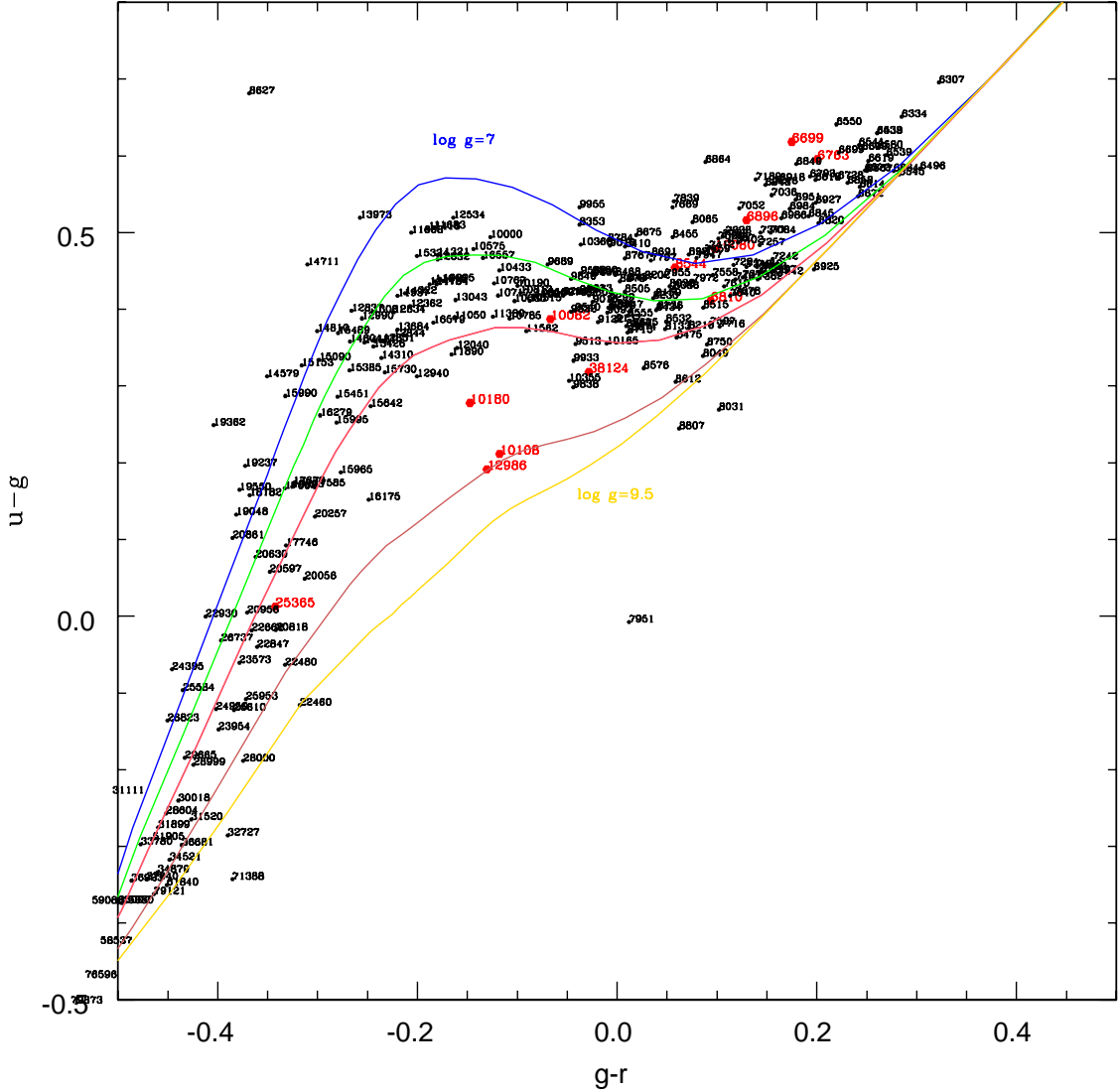}
\caption{Colors for normal DAs (black) and DAHs (red), showing both follow the characteristic
curves, but with an excess of DAHs (red) for higher $\log g$. We again caution the reader that the
$\log g$ estimates are very uncertain for the magnetic stars because the $u$ color,
where the Balmer jump is located,
is heavily affected by the magnetic field.
\label{mean}}
\end{figure}

\begin{figure}
\centering
\includegraphics[width=0.9\textwidth]%
{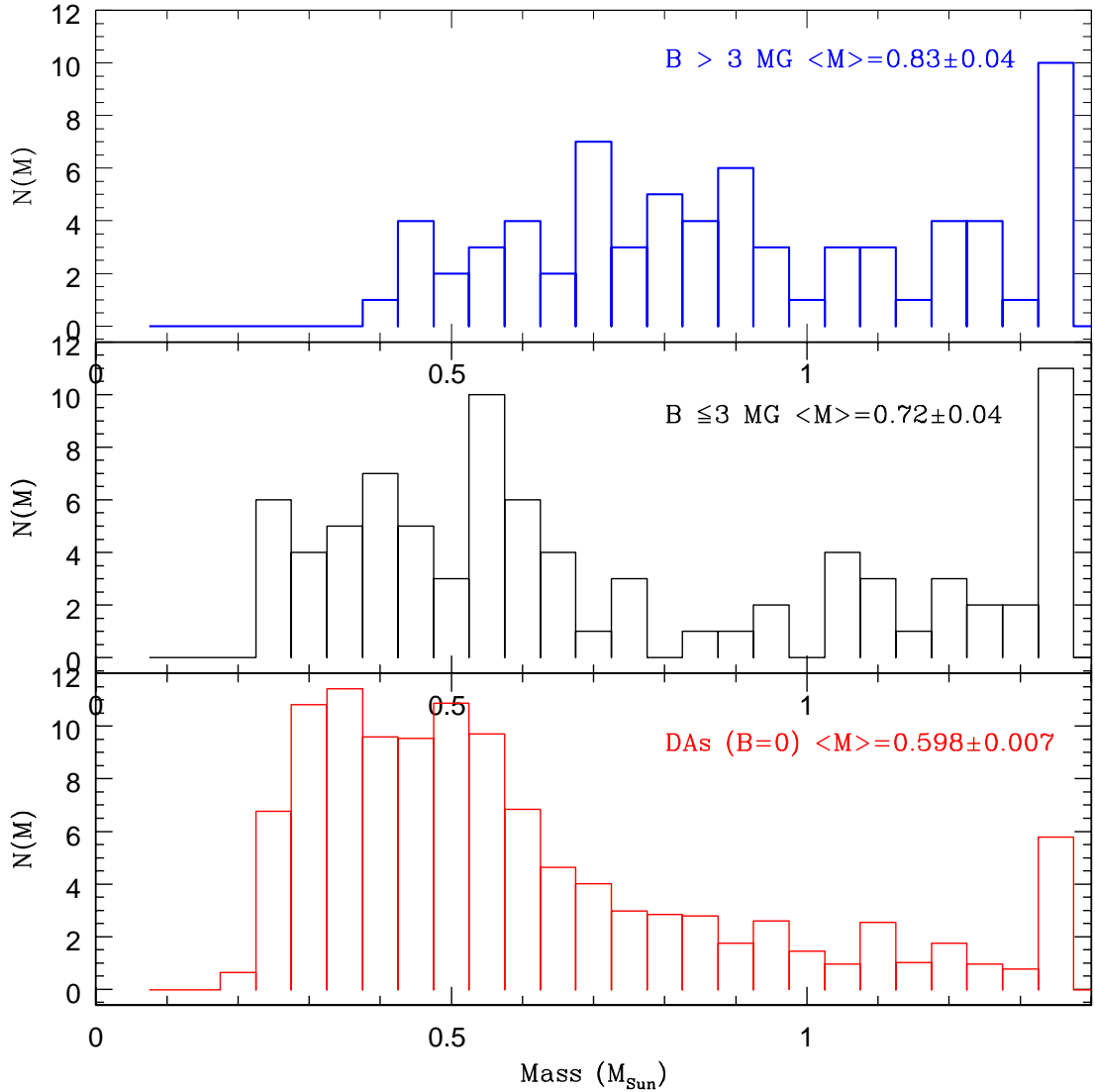}
\caption{Mass histogram from colors for low field (black) and higher field (blue) DAHs and DAs, showing the
apparent  masses increase with increasing magnetic field.}
\label{hist}
\end{figure}

We estimated the masses for every star with
$S/N_g \geq 10$ spectra, 
from their $u,g,r,i,z$ colors, shown in Figure~\ref{mean},
obtaining for the 84 hydrogen-rich magnetic white dwarfs
(DAHs) with $B\leq 3$~MG, $\langle M \rangle=(0.68 \pm 0.04)\, M_\odot$.
For the 71 DAHs with $B>3$~MG, $\langle M \rangle=(0.83 \pm 0.04)\, M_\odot$.
Figure~\ref{hist} shows the mass histogram for stars with fields lower and higher than
$B=3$~MG, compared to nonmagnetic ones, demonstrating there is an increase in the
estimated mass for magnetic stars, but
the estimated mass values are uncertain because the $u$ color is
severely affected by magnetic fields, caused by the $n^4$
dependency of the splittings \citep[e.g.][]{Girven10}.
The estimated masses are much larger than the mean masses for the 1505
bright and hot DA white dwarfs in \citet{dr7}, i.e., those with
$S/N\geq 15$ and $T_{\mathrm{eff}}\geq 13\,000$~K, for which we did
not detect any magnetic field,
${\langle M \rangle}_\mathrm{DA} = (0.593 \pm 0.002)\, M_\odot$. 
Even though we find a higher mass for DAHs,
our mean is much smaller than the mean masses quoted 
for the few previously measured
magnetic DAs. 

\section{Discussion}
Considering only spectra with S/N$\geq$10, we increased the number of known  magnetic DAs by a factor of 2. 
We estimated the field strength for 521 stars.
Our blind test shows we underestimate the number of magnetics
in the simulation and underestimate the field in general. Even for
$S/N\leq 10$, our candidates have at least a 50\% chance of being magnetic,
compared to only 4\% in field white dwarfs.
We showed that the magnetic field changes with time for a few stars we have multiple spectra, pointed that stars
with surface temperatures where convection zone develops seems to show stronger magnetic fields
than hotter stars, and that the mean mass of magnetic stars seems to be on average larger than the mean
mass of non-magnetic stars. 

If the apparent increase in masses shown in Fig.~\ref{mass} were only
caused by magnetic field amplification when the surface convection zone
appears around $T_\mathrm{eff}\leq 13\,000$~K, it should, perhaps,
continue to rise at lower temperatures. But if the mass of the convection
zone becomes high enough that its kinetic energy is of the same order as the
magnetic energy, amplification will not be effective.



\citet{Zorotovic11} estimate the mean WD mass among CVs is 
$\langle M_\mathrm{CV}\rangle = (0.83 \pm 0.23)~M_\odot$, much larger than that found for pre-CVs,
$\langle M_\mathrm{PCV}\rangle = (0.67 \pm 0.21)~M_\odot$,
and single white dwarfs. 
Are all the magnetic white dwarfs descendant of binaries?

The Gaia mission will provide
 parallaxes for all these objects and
 thereby obtain strong constraints on magnetic and convection models and get an
 independent check of the surface gravity ($\log g$) determinations, if we assume a
mass-radius relation.

\section*{Acknowledgments}
S.O. Kepler, J.E.S. Costa,
I. Pelisoli, and V. Pe\-\c{c}a\-nha are sup\-por\-ted by CNPq and FAPERGS-Pronex-Brazil.
BGC is supported by the Austrian Fonds zur F\"{o}rderung der wissenschaftlichen Forschung
 through project P 21830-N16.
BK is supported by the MICINN grant AYA08-1839/ESP, by the ESF EUROCORES
Program EuroGENESIS (MICINN grant EUI2009-04170), by the 2009SGR315 of the
Generalitat de Catalunya and EU-FEDER funds.
D.E. Winget gratefully acknowledge the support of the
US National Science Foundation under grant AST-0909107 and the Norman
Hackerman Advanced Research Program under grant 003658-0252-2009.
A. Kanaan is supported by CNPq.
Funding for the SDSS and SDSS-II was provided by the
Alfred P. Sloan Foundation, the Participating Institutions, the
National Science Foundation, the U.S. Department of Energy, the
National Aeronautics and Space Administration, the Japanese
Monbukagakusho, the Max Planck Society, and the Higher Education
Funding Council for England. The SDSS Web Site is
http://www.sdss.org/.
The SDSS is managed by the Astrophysical Research Consortium
for the Participating Institutions. The Participating Institutions
are the American Museum of Natural History, Astrophysical Institute
Potsdam, University of Basel, University of Cambridge, Case Western
Reserve University, University of Chicago, Drexel University,
Fermilab, the Institute for Advanced Study, the Japan Participation
Group, Johns Hopkins University, the Joint Institute for Nuclear
Astrophysics, the Kavli Institute for Particle Astrophysics and
Cosmology, the Korean Scientist Group, the Chinese Academy of
Sciences (LAMOST), Los Alamos National Laboratory, the
Max-Planck-Institute for Astronomy (MPIA), the Max-Planck-Institute
for Astrophysics (MPA), New Mexico State University, Ohio State
University, University of Pittsburgh, University of Portsmouth,
Princeton University, the United States Naval Observatory, and the
University of Washington.

\begin{table}
\begin{tabular}{lcrcrcrrrr}
\label{dados}
Name (SDSS J)&P-M-F&$B_{H\alpha}$&$B_{H\beta}$&S/N&g&$T_\mathrm{eff}$&$\sigma_T$&log g&$\sigma_g$\cr
&&(MG)&(MG)&&(mag)&(K)&(K)&(cgs)&(cgs)\cr
135141.13+541947.35&1323-52797-293&773&  K&37.71&16.40& 10180&0084&09.64&0.26 \cr
234605.44+385337.69&1883-53271-272&706&  K&17.74&18.89& 99999&0794&05.00&0.01 \cr
100356.32+053825.59&0996-52641-295&668&  K&15.28&18.11& 14019&0140&09.98&0.03 \cr
120609.83+081323.72&1623-53089-573&312&  K&09.34&19.03& 10735&0247&07.65&0.38 \cr
221828.59-000012.21&0374-51791-583&212&  K&17.30&18.13& 12239&0443&09.73&0.24\cr
021148.22+211548.19&2046-53327-048&105&  K&36.09&16.73& 12986&0134&09.99&0.06\cr
085649.68+253441.07&1933-53381-151&90 & -  &28.44&17.51& 10815&0200&09.91&0.10 \cr
080743.33+393829.18&0545-52202-009&65 &  K&04.07&20.14& 10783&0418&08.93&0.67 \cr
170400.01+321328.66&0976-52413-319&56 &  K&04.07&20.42& 25597&1118&09.39&0.21 \cr
023609.38-080823.91&0455-51909-474&54 & -  &06.47&19.75& 08733&0145&09.93&0.08\cr
114006.37+611008.21&0776-52319-042&53 &  K&06.41&19.67& 10540&0326&09.45&0.45\cr
224741.46+145638.76&0740-52263-444&47 &  K&29.37&17.39& 14771&0069&10.00&0.01 \cr
214930.74-072811.97&0644-52173-350&45 &  K&31.93&17.41& 72000&1337&10.00&0.01 \cr
121635.36-002656.22&0288-52000-276&45 &  K&06.51&19.58& 11150&0286&09.70&0.26\cr
160357.93+140929.97&2524-54568-247&43 & -  &16.91&18.29& 10123&0055&09.26&0.44 \cr
101805.04+011123.52&0503-51999-244&40 &  K&49.54&16.31& 10108&0043&09.87&0.11 \cr
094235.02+205208.32&2292-53713-019&38 &  K&14.39&18.44& 13277&0140&09.99&0.02 \cr
160437.36+490809.18&0622-52054-330&38 &  K&20.94&17.90& 10084&0012&09.25&0.60 \cr
125416.01+561204.67&1318-52781-299&37 &  K&08.59&19.01& 10338&0172&08.62&0.53 \cr
151415.66+074446.50&1817-53851-534&36 &  K&14.92&18.84& 10090&0024&09.55&0.38 \cr
082835.82+293448.69&1207-52672-635&35 &  K&06.17&19.74& 13176&0327&09.88&0.13 \cr
114828.99+482731.23&1446-53080-324&33 &  K&17.05&18.16& 89520&5406&10.00&0.01 \cr
075819.57+354443.70&0757-52238-144&32 &  K&22.10&18.20& 12930&0100&10.00&0.01 \cr
080938.10+373053.81&0758-52253-044&31 &  K&10.33&19.01& 11398&0320&09.56&0.29 \cr
142703.35+372110.51&1381-53089-182&30 &34 &23.95&17.55& 49950&0990&10.00&0.01 \cr
172329.14+540755.82&0359-51821-415&30 &  K&10.13&18.80& 10157&0092&09.15&0.50\cr
085153.79+152724.94&2431-53818-238&29 &32 &08.38&19.42& 11300&0388&09.58&0.31 \cr
080440.35+182731.03&2081-53357-442&29 &  K&18.23&18.11& 10135&0063&08.46&0.39 \cr
011423.35+160727.51&2825-54439-548&28 &29  &16.90&19.50& 46537&1332&10.00&0.01 \cr
080359.94+122944.02&2265-53674-033&27 &  K&31.03&17.27& 12347&0109&09.99&0.02 \cr
172932.48+563204.09&0358-51818-239&27 &  K&05.56&20.03& 10277&0182&09.18&0.62\cr
115418.14+011711.41&0515-52051-126&26 &  K&19.69&17.75& 79747&3659&10.00&0.01 \cr
093415.97+294500.43&2914-54533-162&25 &32  &25.07&18.95& 61840&1821&10.00&0.01 \cr
122401.48+415551.91&1452-53112-181&25 &  K&12.32&18.94& 10100&0034&09.35&0.46 \cr
113839.51-014903.00&0327-52294-583&24 &  K&28.51&17.61& 10198&0095&09.47&0.31\cr
215148.31+125525.49&0733-52207-522&22 &  K&20.11&18.10& 43262&1151&09.99&0.02 \cr
232248.22+003900.88&0383-51818-421&22 &  K&08.90&19.12& 13458&0190&09.97&0.04 \cr
023420.63+264801.71&2399-53764-559&21 &  K&28.75&18.39& 57403&0981&10.00&0.01 \cr
105628.49+652313.45&0490-51929-205&21 &  K&09.28&19.71& 43262&1894&09.99&0.02 \cr
125553.39+152555.08&1771-53498-343&20 & 2  &06.74&19.52& 15263&0536&07.68&0.16\cr
125044.43+154957.36*&1770-53171-530&20 &  K&16.23&18.28& 10082&0010&09.66&0.35 \cr
125715.54+341439.38&2006-53476-332&19 &19  &36.15&16.80& 38124&0288&10.00&0.01 \cr
091124.68+420255.85&1200-52668-538&19 &  K&13.77&18.84& 10108&0043&09.51&0.36 \cr
120150.13+614256.93&0778-54525-280& 19&17  &14.68&18.48& 08122&0076&10.00&0.01 \cr
                   &0778-52337-264&  8&  K&13.50\cr
105404.38+593333.34&0561-52295-008&18 &  K&03.51&20.27& 09313&0382&09.93&0.08 \cr
084201.42+153941.89&2429-53799-363&17 &17  &06.90&19.73& 22510&0794&10.00&0.01 \cr
032628.17+052136.35&2339-53729-515&17 &  K&17.94&18.95& 55082&1926&09.99&0.01 \cr
225726.05+075541.71&2310-53710-420&17 &  K&34.35&17.10& 85279&2742&09.99&0.02 \cr
094458.92+453901.15&1202-52672-577&17 &  K&05.79&19.92& 18919&1136&10.00&0.01 \cr
122209.43+001534.06&2568-54153-471&16 &  K&08.33&20.26& 21070&0749&10.00&0.01\cr
                   &0289-51990-349&16&   K&03.23 \cr
053317.32-004321.91&2072-53430-096&16 &11  &17.06&00.00& 71388&5460&07.10&0.24\cr
153829.29+530604.65&0795-52378-637&16 &  K&08.51&19.26& 18116&0547&10.00&0.01 \cr
\end{tabular}
\end{table}
\begin{table}
\begin{tabular}{lcrcrcrrrr}
Name (SDSS J)&P-M-F&$B_{H\alpha}$&$B_{H\beta}$&S/N&g&$T_\mathrm{eff}$&$\sigma_T$&log g&$\sigma_g$\cr
&&(MG)&(MG)&&(mag)&(K)&(K)&(cgs)&(cgs)\cr
131508.97+093713.87&1798-53851-233&14& -  &45.27&16.23& 72414&1249&10.00&0.01\cr
074924.91+171355.45&2729-54419-282&14 &  K&22.57&18.78& 35795&0432&10.00&0.01 \cr
072540.82+321402.12&2695-54409-564&14 &13 &07.43&20.06& 34711&1100&09.99&0.02 \cr
100759.81+162349.64&2585-54097-030&14 &  K&22.19&17.74& 32642&0275&10.00&0.01 \cr
083448.65+821059.00&2549-54523-135&14 &  K&28.14&18.33& 41210&0539&10.00&0.01 \cr
134820.80+381017.25&2014-53460-236&14 &  K&23.58&17.55& 45528&0864&10.00&0.01 \cr
133340.34+640627.38&0603-52056-112&14 &  K&19.01&17.88& 20048&0152&10.00&0.01 \cr
115345.97+133106.61&1762-53415-042&13&  - &05.56&19.81& 12844&0540&09.19&0.40\cr
143235.46+454852.52&2932-54595-542&13 &  K&09.68&19.94& 21053&0492&09.99&0.02 \cr
101428.10+365724.40&1954-53357-393&13 &13 &12.95&18.85& 19580&0342&10.00&0.01 \cr
140716.67+495613.70&1671-53446-453&13 &  K&09.49&19.14& 27376&0610&10.00&0.01 \cr
101428.10+365724.40&1426-52993-021&13 &13 &08.49&18.85& 17225&0724&09.99&0.02 \cr
205233.52-001610.69&0982-52466-019&13 &  K&12.40&18.51& 36761&0831&09.99&0.01 \cr
074947.00+354055.51&0542-51993-639&13 &12 &11.00&19.75& 39100&1213&09.98&0.03\cr
152401.60+185659.21&2794-54537-410&12 &  K&20.77&18.16& 22067&0290&10.00&0.01 \cr
085550.68+824905.20&2549-54523-066&12 &  K&22.40&18.64& 32081&0265&10.00&0.01 \cr
103350.88+204729.40&2376-53770-463&12 &13  &08.00&19.42& 34544&0947&09.98&0.03 \cr
090748.82+353821.5 &1212-52703-187&12 &  K& -   &19.61& 12485&0306&09.85&0.14\cr
001034.95+245131.20&2822-54389-025&11 &11 &10.43&19.84& 19355&0654&10.00&0.01 \cr
075234.95+172524.86&2729-54419-171&11 &11 &28.79&18.46& 39779&0471&10.00&0.01 \cr
                   &1920-53314-106&12 &  K&17.67\cr
202501.11+131025.62&2257-53612-167&11 &  K&23.72&18.77& 28932&0199&10.00&0.01 \cr
120547.48+340811.48&2089-53498-431&11 &12  &09.00&19.63& 23237&0632&10.00&0.01\cr
033145.69+004517.04&0810-52672-391&11 & 11 &45.85&17.20& 28299&0095&10.00&0.01\cr
                   &2049-53350-450&11 &11&33.80\cr
                   &0415-51879-378&12 &13&32.22\cr
                   &0415-51810-370&12 &  K&32.86\cr
081648.71+041223.53&1184-52641-329&11& 10 &03.44&20.39& 12880&0912&09.95&0.06\cr
030407.40-002541.74&0709-52205-120&11& 11 &26.97&17.75& 21828&0227&10.00&0.01\cr
                   &2048-53378-280&10&10&26.18\cr
                   &0411-51817-172&11&10&23.40\cr
                   &0411-51873-172&19&18&22.83\cr
                   &0710-52203-311&11&11&21.25\cr
115917.39+613914.32&0777-52320-069&10 &  K&07.64&18.97& 35770&1193&09.99&0.02 \cr
121209.31+013627.72&0518-52282-285&10 &  K&21.90&18.00& 24706&0182&10.00&0.01 \cr
031824.20+422651.00&2417-53766-568&9.9&9.2 &30.87&18.21& 20274&0098&10.00&0.01 \cr
084008.50+271242.70&1587-52964-059&9.8&10  &08.06&19.17& 19113&0355&09.99&0.02 \cr
172045.37+561214.90&0367-51997-461&9.7&  K&06.24&20.10& 46580&3487&09.95&0.06\cr
153843.11+084238.27&1725-54266-297&9.6&  K&19.58&17.92& 33803&0340&09.99&0.01\cr
165203.68+352815.81&0820-52438-299&9.5&  K&09.99&19.23& 18778&0406&09.94&0.07\cr
034308.18-064127.35&0462-51909-117&9.2&  K&08.82&19.48& 11718&0447&09.57&0.32\cr
153532.25+421305.62&1052-52466-252&9.1&  K&03.21&20.37& 18143&1006&08.10&0.24\cr
091437.35+054453.31&1193-52652-481&8.9&  K&28.18&17.33& 23420&0229&10.00&0.01\cr
123204.20+522548.27&0885-52379-319&8.9&9.6 &10.25&18.82& 08183&0093&09.29&0.10\cr
124851.31-022924.73&2922-54612-607&8.8&8.7 &34.75&18.42& 19835&0070&10.00&0.01\cr
112030.34-115051.14&2874-54561-512&8.8&8.7 &21.88&18.73& 27302&0222&10.00&0.01\cr
093126.14+321946.15&1943-53386-294&8.6&8.3 &07.49&19.23& 16248&0629&09.58&0.12\cr
122249.14+481133.14&1451-53117-582&8.6&8.3 &13.72&18.72& 09790&0093&10.00&0.01\cr
151130.17+422023.00&1291-52735-612&8.4&  K&19.79&17.99& 30882&0265&10.00&0.01\cr
                   &1291-52738-615&12&12 &18.41\cr
154213.48+034800.43&0594-52045-400&8.2&  K&11.35&19.12& 15760&0877&10.00&0.01\cr
113756.50+574022.43&1311-52765-421&8.1&8.3 &29.48&16.87& 10080&0002&09.03&0.80\cr
\end{tabular}
\end{table}
\begin{table}
\begin{tabular}{lcrcrcrrrr}
Name (SDSS J)&P-M-F&$B_{H\alpha}$&$B_{H\beta}$&S/N&g&$T_\mathrm{eff}$&$\sigma_T$&log g&$\sigma_g$\cr
&&(MG)&(MG)&&(mag)&(K)&(K)&(cgs)&(cgs)\cr
100005.67+015859.18&0500-51994-557&8.1&  K&05.03&20.04& 08124&0141&09.96&0.05\cr
171441.08+552711.45&0367-51997-318&8.1&9.2 &06.27&20.23& 36207&1509&07.45&0.25\cr
150813.25+394504.91&1398-53146-633&8.0  &  K&28.17&17.89& 29032&0167&09.52&0.04\cr
074853.08+302543.56&0889-52663-507&8.0  &  K&34.70&17.59& 28932&0116&09.73&0.04\cr
124851.31-022924.69&0337-51997-264&8.0  &  K&12.17&18.42& 18000&0212&09.59&0.05\cr
085523.87+164059.01&2431-53818-522&7.9&  K&12.84&18.56& 28315&0477&10.00&0.01\cr
023445.31+260553.18&2399-53764-487&7.8&7.4 &05.59&20.72& 19105&0558&09.94&0.07\cr
113357.66+515204.85&0879-52365-586&7.7&  K&29.49&17.33& 24938&0173&10.00&0.01\cr
112257.11+322327.80&1979-53431-512&7.5&  K&08.49&19.37& 18771&0482&10.00&0.01\cr
101618.37+040920.58&0574-52355-166&7.2&  K&03.47&20.29& 12275&0732&09.93&0.07\cr
130033.48+590407.05&2461-54570-015&6.8&6.5 &26.57&18.23& 06300&0004&10.00&0.01\cr
014230.57+003502.66&1907-53315-427&6.6&5.7 &09.96&19.99& 10676&0131&07.90&0.13\cr
                   &1907-53265-425&1.7&  - &09.96\cr
080502.29+215320.54&1584-52943-132&6.6&7.0 &18.52&18.61& 37141&0575&09.39&0.07\cr
083438.29+153817.48&2427-53815-321&6.5&5.7 &03.40&20.24& 12290&0569&08.27&0.24\cr
122748.85+385546.34&1986-53475-090&6.5&7.0 &06.24&19.49& 18116&0554&09.81&0.12\cr
143019.06+281100.87&2134-53876-423&6.2&  K&24.73&17.68& 07906&0053&09.48&0.04\cr
111010.50+600141.44&2414-54526-323&6.1&6.5 &31.79&17.99& 35974&0282&09.57&0.04\cr
                   &0950-52378-568&6.7&  K&17.38\cr
151745.19+610543.59&0613-52345-446&6.1&  K&03.68&20.52& 13748&1001&09.92&0.09\cr
231951.73+010909.32&0382-51816-565&6.1&  K&14.00&18.51& 07525&0062&09.02&0.11\cr
170916.37+234111.33&1688-53462-508&6.0&  - &04.04&20.29& 17102&0998&09.62&0.17 \cr
090632.66+080715.96&1300-52973-148&6.0&  K&17.54&18.66& 46580&1335&09.93&0.07\cr
102220.70+272539.85&2350-53765-543&5.8&  K&05.32&20.06& 14780&0667&09.12&0.17\cr
165249.09+333444.90&1175-52791-095&5.8&  K&09.89&18.64& 09593&0102&09.32&0.10\cr
171831.24+280825.72&0980-52431-434&5.6&  - &05.35&19.87& 44249&3458&07.15&0.46 \cr
153315.26+564200.32&0614-53437-579&5.6&6.1 &04.20&20.39& 08017&0164&09.63&0.28\cr
144614.00+590216.73&0608-52081-140&5.6&7.0 &04.47&20.11& 15794&0569&09.11&0.17\cr
100657.52+303338.10&1953-53358-415&5.5&4.8 &11.90&18.82& 14462&0303&08.94&0.08\cr
215248.44-010324.17&1107-52968-374&5.5&  K&04.67&20.29& 08990&0208&08.20&0.35\cr
111150.80+663736.20&0490-51929-639& - &5.4 &05.23&20.05& 18084&0681&07.59&0.16\cr
014313.18+231524.60&2064-53341-024&5.3&  - &04.40&20.73& 43788&3957&08.91&0.49 \cr
134043.10+654349.20&0497-51989-182&5.3&  K&15.47&18.47& 19819&0249&09.99&0.02\cr
134043.11+654349.26&2460-54616-013&5.1&5.2 &22.77&18.47& 18612&0179&09.51&0.03\cr
083627.35+154850.31&2276-53712-107&5.1&7.0 &09.49&19.28& 47547&2422&08.84&0.22\cr
013533.19+132249.90&0426-51882-291&5.0&5.2 &02.14&20.46& 13480&1164&08.70&0.37\cr
125434.65+371000.19&1989-53772-041&4.9&  K&46.84&16.01& 25365&0140&09.27&0.02\cr
053126.76+001738.13&2072-53430-427&4.6&3.5 &05.35&20.69& 17753&0907&07.70&0.17 \cr
012339.94+405241.88&2063-53359-403&4.6&3.5 &06.41&20.27& 36761&1448&07.48&0.22 \cr
120806.26+144942.97&1764-53467-112&4.6&2.5 &04.03&20.16& 11691&0507&08.27&0.26 \cr
144934.73+025502.62&0537-52027-566&4.6&4.4 &09.86&18.98& 07961&0097&09.36&0.16\cr
144408.76+024327.75&0537-52027-371&4.6&3.9 &06.58&19.45& 10129&0063&09.22&0.54\cr
075816.64+121428.68&2265-53674-199&4.5&3.9 &09.47&19.00& 13625&0354&08.39&0.09\cr
\end{tabular}
\end{table}
\begin{table}
\begin{tabular}{lcrcrcrrrr}
Name (SDSS J)&P-M-F&$B_{H\alpha}$&$B_{H\beta}$&S/N&g&$T_\mathrm{eff}$&$\sigma_T$&log g&$\sigma_g$\cr
&&(MG)&(MG)&&(mag)&(K)&(K)&(cgs)&(cgs)\cr
064828.76+840340.87&2548-54152-616&4.3&4.4 &11.34&19.75& 17290&0320&09.24&0.07\cr
102535.40+282034.79&2351-53772-007&4.3&3.9 &07.47&19.74& 24023&0790&07.85&0.14\cr
090522.06+205736.32&2284-53708-091&4.3&4.8 &08.34&19.36& 12469&0389&08.17&0.14\cr
004528.87+004616.44&1905-53706-526&4.3&3.0 &16.40&19.26& 17439&1340&09.27&0.07 \cr
153057.50+394615.03&1293-52765-613&4.3&  - &03.45&20.38& 15575&0794&07.32&0.22 \cr
093447.90+503312.19&0901-52641-373&4.3&  K&10.97&18.82& 08946&0098&09.99&0.01\cr
105328.14+505155.14&0876-52669-510&4.3&4.4 &06.01&19.87& 09055&0148&09.07&0.20\cr
003111.75+134919.53&0417-51821-084&4.3&4.8 &03.14&20.21& 16953&1101&08.49&0.22\cr
031323.65-001659.86&0413-51929-313&4.3&4.4 &06.26&19.79& 65455&8331&08.68&0.56\cr
101339.47+352639.61&1954-53357-250&4.2&1.5 &05.20&20.02& 16128&0547&07.53&0.15 \cr
124836.32+294231.26&2457-54180-112&4.0&  K&25.85&17.82& 06489&0020&10.00&0.01\cr
084510.23+112405.61&2428-53801-089&4.0&4.8 &04.07&20.01& 13659&1059&07.22&0.25\cr
152203.45+203438.96&2156-54525-031&4.0&3.9 &11.06&18.79& 19850&0453&07.85&0.08\cr
124816.82+411051.23&1985-53431-445&4.0&3.5 &09.71&19.14& 20265&0440&09.49&0.08\cr
154305.67+343223.68&1402-52872-145&4.0&  K&13.55&18.33& 30620&0335&09.24&0.07\cr
091314.46+410838.75&1200-52668-079&4.0&2.0 &04.47&20.20& 12110&0599&08.37&0.20 \cr
144345.84+482008.97&1047-52733-099&4.0&4.0 &05.00&19.75& 16811&0606&07.43&0.16 \cr
085130.57+353117.59&0934-52672-080&4.0&4.0 &04.99&19.79& 07581&0119&08.98&0.50 \cr
104356.64+650057.24&0489-51930-079&4.0&3.9 &05.26&20.03& 12488&0399&07.84&0.19\cr
091611.08+124808.09&2577-54086-427&3.9&3.9 &06.15&19.89& 15711&0465&07.78&0.12\cr
102239.06+194904.33&2374-53765-544&3.9&3.5 &11.42&19.02& 10060&0096&09.42&0.08\cr
165029.91+341125.50&1175-52791-482&3.9&  K&11.26&18.77& 11921&0186&09.25&0.09\cr
121110.30+203429.05&2609-54476-564&3.8&3.0 &08.72&19.77& 23497&0635&07.78&0.10\cr
114917.23+300016.08&2222-53799-593&3.8&3.5 &15.53&18.77& 09927&0066&08.69&0.07\cr
224742.58-003317.39&0676-52178-319&3.8&4.4 &05.52&19.63& 57600&7109&07.21&0.62\cr
132203.94+193223.08&2619-54506-423&3.6&2.2 &04.61&20.30& 11413&0351&07.91&0.20\cr
093726.27+205756.98&2361-53762-408&3.6&3.5 &05.25&19.99& 12266&0522&08.14&0.20\cr
092355.98+243552.86&2291-53714-552&3.6&3.9 &03.99&20.41& 12976&0662&08.57&0.21\cr
113048.38+305720.74&1974-53430-222&3.6&3.9 &07.09&19.65& 12647&0425&09.66&0.15\cr
211125.84+110219.69&1890-53237-390&3.6&3.5 &25.85&17.40& 18577&0123&09.45&0.02\cr
235431.38+365019.13&1881-53261-512&3.6&3.5 &06.44&20.48& 14078&1251&09.29&0.17 \cr
124806.38+410427.15&1456-53115-190&3.6&  K&19.35&17.90& 06574&0032&10.00&0.01\cr
120728.96+440731.62&1369-53089-048&3.6&3.8 &06.87&19.32& 18813&0520&07.97&0.12\cr
094711.11+003754.69&0480-51989-082&3.6&4.4 &07.56&19.74& 13827&3547&08.11&0.33 \cr
081136.34+461156.44&0439-51877-523&3.6&4.8 &06.29&19.82& 27097&1515&07.20&0.37\cr
081523.35+084346.43&2571-54055-256&3.5&3.5 &04.16&20.27& 13816&0655&08.11&0.20\cr
203256.48+142651.99&2258-54328-278&3.5&2.6 &07.85&20.57& 19968&0886&07.92&0.16\cr
114833.25+303921.27&2222-53799-577&3.5&3.5 &04.71&20.23& 11053&0301&08.62&0.17\cr
090937.95+250820.64&2086-53401-582&3.5&3.7 &09.80&19.01& 11012&0154&08.80&0.09\cr
\end{tabular}
\end{table}
\begin{table}
\begin{tabular}{lcrcrcrrrr}
Name (SDSS J)&P-M-F&$B_{H\alpha}$&$B_{H\beta}$&S/N&g&$T_\mathrm{eff}$&$\sigma_T$&log g&$\sigma_g$\cr
&&(MG)&(MG)&&(mag)&(K)&(K)&(cgs)&(cgs)\cr
153301.50+550840.72&0614-53437-079&3.5&2.6 &05.72&19.94& 09644&0153&08.23&0.24\cr
154135.02+030051.14&0594-52045-253&3.5&4.4 &04.15&20.33& 11024&0366&09.18&0.26\cr
010225.14-005458.15&0693-52254-099& - &3.5 &07.95&19.26& 18564&1798&09.00&0.14\cr
081716.39+200834.89&2082-53358-444&3.4&  K&16.66&18.35& 06886&0043&09.99&0.02\cr
011130.67+141049.67&2825-54439-089&3.3&3.0 &06.99&19.97& 15022&0784&07.91&0.13\cr
091340.19+114112.41&2576-54086-029&3.3&3.9 &06.80&19.54& 15422&0486&07.76&0.12\cr
212329.46-081004.44&2320-54653-090&3.3&3.0 &08.36&19.98& 16056&0373&08.88&0.10\cr
131544.04+262333.30&2243-53794-170&3.3&3.0 &06.30&19.69& 09008&0151&09.26&0.17\cr
163600.25+354625.33&2185-53532-320&3.3&3.9 &18.32&19.21& 24080&0368&07.99&0.05\cr
083918.12+212143.79&2084-53360-100&3.3&3.9 &04.11&20.10& 10813&0394&09.82&0.15\cr
083041.78+204233.84&2083-53359-553&3.3&3.5 &07.92&19.39& 16175&0467&08.37&0.09\cr
125101.98+351913.59&1987-53765-570&3.3&2.6 &09.07&19.07& 19827&0413&08.34&0.09\cr
074403.81+495440.02&1868-53318-410&3.3&1.5 &14.08&18.75& 36390&0736&09.65&0.12 \cr
161047.77+235301.87&1852-53534-567&3.3&2.6 &04.95&19.83& 09365&0167&09.95&0.06\cr
114827.96+153356.97&1762-53415-372&3.3&3.0 &04.97&20.17& 11515&0418&09.07&0.16 \cr
155436.25+413956.61&1334-52764-235&3.3&  - &03.69&20.12& 19749&0960&09.24&0.22 \cr
110735.32+085924.59&1221-52751-177&3.3&3.5 &13.05&18.43& 18997&0344&09.15&0.06 \cr
082533.22+412400.15&0760-52264-640&3.3&3.5 &05.43&19.79& 16149&0590&07.78&0.14\cr
224103.30+132853.25&0739-52520-143&3.3&3.5 &05.76&19.78& 22940&0856&07.36&0.19\cr
094815.27+041648.67&0570-52266-632&3.3&3.5 &04.44&19.75& 12311&0704&07.64&0.23\cr
095211.45+563020.74&0557-52253-170&3.3&3.5 &11.81&18.69& 22652&0509&07.98&0.08\cr
084937.77+561949.27&0483-51902-296&3.3&4.4 &05.27&20.11& 10008&0195&09.04&0.25\cr
094012.89-000009.88&0476-52314-597&3.3&2.6 &05.98&19.63& 22232&0843&08.07&0.14\cr
094046.29+595415.92&0453-51915-325&3.3&3.0 &07.05&19.57& 10769&0228&07.83&0.17\cr
121105.25-004628.45&0287-52023-253&3.3&4.4 &13.99&18.69& 22490&0371&09.06&0.06\cr
115817.44+331719.26&2095-53474-337&3.2&2.6 &05.15&19.80& 19976&0697&08.31&0.15\cr
115554.95+083549.54&1622-53385-447&3.2&2.6 &12.38&18.88& 09888&0089&09.37&0.07\cr
120125.40+084800.44&1228-52728-220&3.2&  - &09.39&18.68& 06965&0064&09.99&0.02\cr
095738.55+194601.95&2363-53763-097&3.1&3.5 &04.01&20.21& 13411&0760&09.44&0.24\cr
230758.83+000500.32&0678-52884-498&3.1&3.4 &05.12&20.04& 13923&1877&08.02&0.30\cr
145801.10+040917.03&0588-52029-639&3.1&3.5 &03.97&19.90& 08729&0133&05.15&0.15\cr
092527.47+011328.66&0475-51965-315&3.1&  K&12.04&18.60& 11016&0104&09.01&0.10\cr
032137.44+010437.38&0413-51821-578&3.1&3.5 &03.76&19.92& 13729&1414&08.07&0.23\cr
025837.20+000019.27&0410-51877-065&3.1&4.4 &05.38&19.73& 11053&0688&08.47&0.21\cr
154141.85+173026.25&2795-54563-603&3.0&2.2 &06.13&19.86& 14074&0605&07.95&0.14\cr
121211.28+185228.97&2609-54476-005&3.0&3.0 &06.73&19.92& 20758&0675&08.10&0.13\cr
103002.67+163927.45&2592-54178-417&3.0&2.6 &04.68&20.09& 18334&0747&07.99&0.17\cr
083310.57+234812.72&2330-53738-109&3.0&2.2 &06.08&20.48& 11111&0250&08.80&0.21\cr
012115.45+321010.04&2061-53711-506&3  &5.0 &06.88&20.36& 17585&0583&08.03&0.12 \cr
131050.04+143520.47&1772-53089-559&3.0&  - &05.23&19.74& 69903&1018& 7.91&0.58\cr
151102.75+433558.73&1677-53148-151&3.0&2.6 &04.19&20.36& 09472&0194&08.02&0.30\cr
093409.91+392759.33&1215-52725-241&3.0&2.6 &15.62&18.35& 10985&0091&09.03&0.07\cr
105216.00+530120.32&1010-52649-597&3.0&  - &04.78&20.25& 12698&0633&07.98&0.20 \cr
100822.39+015307.46&0501-52235-077&3.0&3.8 &08.31&19.50& 57534&6042&07.27&0.39\cr
090343.14+011846.39&0470-51929-403&3.0&2.1 &06.34&19.62& 07930&0090&05.10&0.11\cr
\end{tabular}
\end{table}
\begin{table}
\begin{tabular}{lcrcrcrrrr}
Name (SDSS J)&P-M-F&$B_{H\alpha}$&$B_{H\beta}$&S/N&g&$T_\mathrm{eff}$&$\sigma_T$&log g&$\sigma_g$\cr
&&(MG)&(MG)&&(mag)&(K)&(K)&(cgs)&(cgs)\cr
084155.74+022350.56&0564-52224-248&2.9&  K&16.21&18.99& 06587&0038&10.00&0.01\cr
161929.63+131833.45&2530-53881-332&2.8&2.6 &05.73&19.70& 16471&0554&07.74&0.15\cr
091305.88+173932.93&2439-53795-357&2.8&3.0 &05.56&19.80& 12792&0565&07.88&0.16\cr
114529.27+300824.36&2222-53799-483&2.8&3.0 &07.75&19.62& 19811&0483&07.77&0.14\cr
113215.39+280934.31&2219-53816-329&2.8&  K&29.39&17.02& 06896&0029&10.00&0.01\cr
161147.94+211136.64&2205-53793-542&2.8&3.5 &06.96&20.05& 10303&0190&07.97&0.17\cr
154856.94+230727.92&2169-53556-491&2.8&2.6 &13.73&18.73& 09732&0075&08.62&0.09\cr
012105.53+393239.65&2062-53381-337&2.8&3.0 &10.67&19.67& 21640&0511&07.99&0.08\cr
235318.57+380913.17&1883-53271-050&2.8&1.0 &10.54&19.44& 13142&0679&09.48&0.09 \cr
153554.18+404414.16&1679-53149-109&2.8&1.7 &04.31&20.27& 21818&1067&07.72&0.21\cr
125040.82+590341.61&2461-54570-219&2.7&2.6 &05.50&20.01& 22340&0802&09.00&0.16\cr
011739.82+242236.26&2060-53706-086&2.7&2.0 &10.36&19.54& 14558&0391&07.80&0.09 \cr
112014.62+400422.61&1980-53433-630&2.7&2.0 &08.99&19.50& 13214&0461&09.00&0.10 \cr
101834.81+303330.46&1956-53437-197&2.7&  - &05.75&19.78& 30657&0876&09.97&0.04 \cr
100715.55+123709.51&1745-53061-313&2.7&4.0 &09.27&18.76& 19772&0388&09.69&0.08\cr
143308.50+102623.05&1709-53533-511&2.7&1.5 &17.07&18.28& 06657&0033&10.00&0.01 \cr
010214.99+462620.93&1472-52913-278&2.7&1.5 &12.20&00.00& 08033&0068&07.42&0.17 \cr
111812.67+095241.36&1222-52763-477&2.7&  K&10.86&18.75& 14306&0333&08.90&0.08\cr
084910.13+044528.71&1188-52650-635&2.7&1.6 &08.11&19.29& 09959&0155&05.99&0.18 \cr
220524.61+010503.35&1105-52937-404&2.7&2.5 &09.07&19.46& 16385&0511&09.19&0.08 \cr
095919.86+573542.89&0558-52317-158&2.7&2.6 &03.88&20.27& 12278&1478&07.93&0.47\cr
151436.66+152058.48&2766-54242-497&2.6&3.0 &14.98&18.53& 11318&0104&09.25&0.06\cr
225338.68+301803.47&2627-54379-021&2.6&2.2 &05.44&19.86& 16930&0512&08.32&0.13\cr
161050.03+094302.45&2526-54582-115&2.6&2.6 &04.88&19.97& 25289&1193&07.78&0.19\cr
160219.42+112606.53&2525-54569-303&2.6&2.6 &05.62&19.85& 09800&0151&08.27&0.20\cr
030432.89+365537.88&2443-54082-137&2.6&2.6 &05.07&20.83& 19718&1180&07.73&0.21\cr
084219.70+122128.36&2428-53801-221&2.6&2.6 &04.79&19.74& 11929&0765&07.74&0.22\cr
103532.53+212603.56&2376-53770-534&2.6&2.7 &23.44&17.40& 06763&0031&09.99&0.02\cr
092041.54+221545.53&2319-53763-557&2.6&2.6 &06.73&20.35& 17734&0574&07.99&0.12\cr
080210.39+153033.83&2266-53679-534&2.6&2.2 &08.79&19.79& 11364&0205&08.07&0.12\cr
151606.35+274647.04&2154-54539-637&2.6&3.0 &14.23&18.38& 41076&1071&07.94&0.11\cr
141906.20+254356.51&2131-53819-317&2.6&  K&26.50&17.41& 09906&0033&08.62&0.04\cr
074958.59+161120.32&2080-53350-349&2.6&2.6 &04.28&20.25& 14499&0638&07.54&0.17\cr
073953.18+204900.26&2079-53379-051&2.6&3.0 &08.80&19.92& 15385&0413&07.78&0.09\cr
013909.15+230845.00&2064-53341-163&2.6&  - &05.03&20.69& 19303&0850&08.12&0.18\cr
121529.85+335158.62&1999-53503-238&2.6&2.8 &08.68&19.17& 17847&0373&08.07&0.09\cr
135654.77+343617.25&1838-53467-240&2.6&2.0 &06.95&19.91& 34146&0949&07.82&0.19\cr
132002.48+131901.57&1773-53112-011&2.6&  K&06.72&19.50& 15204&0449&07.87&0.11\cr
081144.60+071030.58&1756-53080-585&2.6&2.6 &08.64&19.41& 17470&0375&07.64&0.10\cr
170019.42+245701.18&1693-53446-169&2.6&2.0 &04.44&20.00& 29422&0996&07.89&0.21\cr
134017.71+594552.42&0786-52319-254&2.6&3.0 &12.17&19.03& 35770&0685&07.57&0.12\cr
032300.93+002221.32&0712-52199-427&2.6&2.6 &03.93&20.43& 11906&1209&07.51&0.35\cr
004248.19+001955.26&0690-52261-594&2.6&  K&08.16&19.52& 10398&0135&09.98&0.03\cr
233328.30-002036.70&0682-52525-317&2.6&2.7 &05.05&19.98& 08401&0169&08.18&0.31\cr
225230.63+003232.68&0676-52178-481&2.6&3.0 &06.21&19.65& 61017&6944&08.14&0.38\cr
204626.15-071036.98&0635-52145-227&2.6&  K&20.70&17.95& 09385&0049&09.19&0.06\cr
220823.66-011534.05&0373-51788-086&2.6&2.3 &14.46&21.75& 29770&0369&07.36&0.08\cr
161425.46+493244.91&2884-54526-254&2.5&3.0 &07.64&19.64& 13744&0570&07.44&0.12\cr
100645.00+144250.28&2586-54169-479&2.5&2.6 &08.48&19.64& 11436&0247&07.95&0.13\cr
082939.25+100937.73&2572-54056-399&2.5&2.2 &08.13&19.72& 15613&0489&08.15&0.09\cr
085106.13+120157.84&2430-53815-229&2.5&  K&28.17&16.98& 14078&0110&08.45&0.03\cr
091833.32+205536.97&2319-53763-209&2.5&2.6 &22.03&18.41& 15584&0173&08.46&0.03\cr
                   &2290-53727-247&2.3&2.4 &16.41\cr
                   &2288-53699-547&2.7& -  &13.15\cr
080638.51+075647.55&2076-53442-521&2.5&2.6 &09.66&19.98& 15580&0409&07.92&0.09\cr
\end{tabular}
\end{table}
\begin{table}
\begin{tabular}{lcrcrcrrrr}
Name (SDSS J)&P-M-F&$B_{H\alpha}$&$B_{H\beta}$&S/N&g&$T_\mathrm{eff}$&$\sigma_T$&log g&$\sigma_g$\cr
&&(MG)&(MG)&&(mag)&(K)&(K)&(cgs)&(cgs)\cr
073001.65+362713.10&2073-53728-482&2.5&2.6 &10.17&19.65& 13877&0554&08.13&0.09\cr
103935.51+295413.59&1969-53383-215&2.5&3.0 &12.77&18.62& 14388&0357&09.08&0.07 \cr
235503.84+350659.75&1881-53261-042&2.5&2.2 &08.17&19.51& 28094&0767&07.78&0.15\cr
073741.50+470421.09&1867-53317-457&2.5&2.6 &13.61&18.83& 56439&2781&07.65&0.19\cr
133828.43+415943.85&1464-53091-605&2.5&3.5 &09.93&18.92& 09270&0094&09.95&0.06\cr
093356.40+102215.69&1303-53050-525&2.5&  K&11.68&18.77& 09075&0082&08.81&0.10\cr
104113.70+083505.23&1240-52734-105&2.5&  - &04.81&19.84& 62299&8856&09.82&0.18 \cr
081130.21+305720.54&0861-52318-096&2.5&3.5 &07.35&19.63& 08045&0101&09.87&0.13\cr
105709.82+041130.37&0580-52368-274&2.5&  K&23.80&17.70& 07287&0035&08.65&0.05\cr
034511.11+003444.27&0416-51811-590&2.5&2.6 &16.89&18.65& 07157&0065&09.30&0.12\cr
220435.05+001242.95&0372-52173-626&2.5&  K&06.11&19.35& 10426&0183&08.98&0.22\cr
173915.64+545059.26&0360-51816-547&2.5&2.2 &05.27&19.84& 12104&0799&07.70&0.23\cr
083420.29+131759.52&2426-53795-387&2.4&3.5 &05.05&19.84& 14114&0644&08.01&0.15\cr
123414.11+124829.58&1616-53169-423&2.4&  K&26.06&17.39& 08544&0049&09.27&0.04\cr
164357.02+240201.31&1414-53135-191&2.4&  K&07.44&19.24& 18715&0521&07.99&0.10\cr
112926.23+493931.86&0966-52642-474&2.4&  K&20.09&18.02& 20160&0210&09.33&0.04\cr
093921.25+581421.45&0452-51911-553&2.4&3.0 &04.62&20.02& 12776&0447&05.17&0.15\cr
154524.79+010127.54&2955-54562-061&2.3&2.2 &04.38&19.92& 13074&1045&07.81&0.23\cr
064532.74+280330.46&2694-54199-201&2.3&3.0 &10.29&19.75& 11351&1245&08.33&0.19\cr
122100.20+244443.75&2657-54502-026&2.3&2.2 &09.03&19.28& 11165&0176&07.81&0.12\cr
173208.55+631950.33&2561-54597-087&2.3&2.2 &09.15&20.05& 16082&0457&08.02&0.10\cr
090907.15+193840.65&2286-53700-176&2.3&2.6 &05.22&19.99& 09871&0176&08.04&0.20\cr
205000.94+170145.59&2259-53565-530&2.3&2.6 &08.24&19.80& 15167&0388&07.98&0.09\cr
113055.05+260115.62&2218-53816-081&2.3&2.6 &05.01&20.07& 15493&0576&07.82&0.16\cr
141813.22+312340.13&2129-54243-426&2.3&2.2 &09.39&19.13& 07507&0076&09.90&0.10\cr
                   &2129-54252-426&2.3&2.2 &09.39\cr
091305.58+260748.62&2087-53415-383&2.3&2.6 &04.01&20.28& 12200&0520&08.17&0.20\cr
013314.21+235247.30&2064-53341-343&2.3&  - &11.09&19.58& 99018&3502&05.25&0.94\cr
120924.84+331716.41&2004-53737-418&2.3&  - &10.62&19.33& 06582&0059&09.90&0.12 \cr
081632.26+522645.27&1781-53297-148&2.3&1.5 &08.27&19.29& 07051&0085&09.50&0.23 \cr
164626.65+222645.45&1570-53149-615&2.3&2.2 &09.31&19.15& 14991&0441&08.40&0.08\cr
114852.78+430753.14&1447-53120-323&2.3&  - &06.03&19.69& 53961&5373&07.74&0.42 \cr
091220.26+075537.71&1301-52976-241&2.3&3.0 &05.05&20.20& 12920&0556&09.26&0.16 \cr
082239.55+304857.26&0931-52619-078&2.3&2.0 &04.64&20.33& 14347&0666&06.48&0.21 \cr
030417.84-003216.31&0709-52205-107&2.3&3.0 &05.63&19.90& 10971&1015&08.09&0.32\cr
094351.26+010104.13&0480-51989-251&2.3&3.5 &05.94&20.13& 09081&0146&07.81&0.32\cr
092242.42+011422.80&0473-51929-003&2.3&3.0 &06.73&19.65& 31980&0771&07.72&0.25\cr
034240.64-073504.00&0462-51909-084&2.3&3.5 &04.97&20.14& 14621&1586&08.03&0.23\cr
021230.01+122557.17&0428-51883-046&2.3&1.7 &04.53&20.24& 14334&2878&07.62&0.30\cr
013920.55+152218.81&0426-51882-524&2.3&3.3 &04.49&19.86& 13386&0857&08.49&0.31\cr
231432.88-011320.52&0382-51816-289&2.3&2.6 &03.38&20.43& 11787&0885&09.69&0.22\cr
112852.88-010540.82&0326-52375-565&2.3&  K&03.56&20.38& 14189&1254&07.78&0.25\cr
141309.30+191832.01&2772-54529-217&2.2&1.9 &21.46&18.20& 18625&0186&09.33&0.03\cr
102746.58+435156.30&2567-54179-306&2.2&2.2 &08.11&19.73& 14462&0452&08.35&0.10\cr
154550.72+132040.19&2517-54567-065&2.2&2.2 &08.46&19.32& 12682&0501&08.06&0.14\cr
030913.96+373057.90&2443-54082-030&2.2&3.0 &04.40&20.83& 24115&1662&07.44&0.24\cr
031929.02+410316.92&2417-53766-064&2.2&1.3 &07.20&20.20& 21275&0844&07.93&0.13\cr
023542.73+241653.78&2399-53764-030&2.2&2.6 &08.70&19.90& 12472&0515&08.13&0.13\cr
024903.02+332737.09&2398-53768-313&2.2&2.2 &08.92&19.82& 18495&0502&07.70&0.10\cr
102429.18+281435.40&2351-53786-001&2.2&2.2 &10.05&19.09& 16884&0345&07.43&0.08\cr
100727.33+281457.81&2348-53757-279&2.2&2.2 &05.38&19.83& 10783&0242&07.99&0.29\cr
212425.74-064837.14&2320-54653-544&2.2&2.2 &13.64&19.51& 16822&0399&08.98&0.06\cr
083945.56+200015.76&2277-53705-484&2.2&  K&22.28&17.84& 17326&0143&09.02&0.03\cr
082817.61+181752.64&2275-53709-298&2.2&2.6 &09.03&19.53& 13673&0412&08.09&0.09\cr
105833.57+372401.35&2091-53447-464&2.2&2.2 &06.76&19.44& 11457&0266&08.79&0.14\cr
012815.72+391130.02&2063-53359-063&2.2&2.6 &05.16&20.29& 15153&0624&07.90&0.14\cr
224707.32+010058.47&1901-53261-564&2.2&2.2 &06.72&20.43& 20622&0717&08.32&0.14\cr
151625.08+280320.92&1846-54173-280&2.2&2.2 &35.41&16.62& 06810&0022&10.00&0.00\cr
145415.01+432149.51&1290-52734-469&2.2&  K&08.42&19.07& 14437&0302&08.62&0.09\cr
101420.38+060254.02&0996-52641-025&2.2&2.0 &03.96&19.84& 13023&0515&07.56&0.18\cr
085830.85+412635.12&0830-52293-070&2.2&2.2 &33.00&17.06& 06699&0021&10.00&0.00\cr
\end{tabular}
\end{table}
\begin{table}
\begin{tabular}{lcrcrcrrrr}
Name (SDSS J)&P-M-F&$B_{H\alpha}$&$B_{H\beta}$&S/N&g&$T_\mathrm{eff}$&$\sigma_T$&log g&$\sigma_g$\cr
&&(MG)&(MG)&&(mag)&(K)&(K)&(cgs)&(cgs)\cr
164703.24+370910.29&0818-52395-026&2.2&  K&20.88&17.63& 17872&0177&09.00&0.03\cr
234623.69-102357.03&0648-52559-142&2.2&  K&14.87&18.43& 07447&0048&10.00&0.01\cr
130807.48-010117.05&0294-51986-089&2.2&2.8 &09.76&19.13& 11699&0253&08.59&0.10\cr
084936.81+224754.97&2085-53379-131&2.1&2.1 &05.19&19.72& 18694&0659&08.07&0.14\cr
120803.24+625815.33&0778-54525-511&2.1&1.7 &06.72&19.76& 62454&7120&07.46&0.46\cr
004122.49-110432.49&0655-52162-091&2.1&3.5 &07.04&19.26& 06977&0077&09.94&0.07\cr
150220.91+001721.08&0539-52017-202&2.1&2.8 &03.85&19.86& 07601&0175&08.63&0.54\cr
003232.07+153126.55&0418-51817-346&2.1&2.4 &05.01&19.77& 12616&0597&07.99&0.17\cr
153349.02+005916.15&2954-54561-048&2.0&2.6 &16.54&18.18& 14920&0181&08.40&0.04\cr
112216.04-122250.99&2874-54561-071&2.0&1.9 &07.02&20.42& 15259&0887&08.16&0.14\cr
173056.42+433000.41&2820-54599-185&2.0&2.4 &05.35&20.29& 12993&0616&08.15&0.20\cr
164649.56+120547.10&2817-54627-308&2.0&2.2 &13.48&19.13& 80899&6363&07.74&0.26\cr
224854.52+303845.57&2627-54379-152&2.0&2.2 &06.30&19.85& 18855&0677&07.82&0.14\cr
083801.81+092548.30&2573-54061-246&2.0&2.6 &05.28&20.16& 22540&1028&07.07&0.24\cr
174235.19+640028.36&2561-54597-021&2.0&1.7 &05.44&20.37& 17004&0832&07.88&0.18\cr
105556.92+483652.49&2410-54087-494&2.0&2.2 &07.31&20.31& 13559&0781&08.50&0.15\cr
092524.26+175712.64&2360-53728-217&2.0&1.9 &10.41&18.80& 17314&0309&07.90&0.08\cr
101059.22+284359.73&2348-53757-485&2.0&2.1 &04.07&20.33& 14512&2382&08.08&0.46\cr
030522.15+050213.33&2322-53727-024&2.0&2.6 &07.15&20.20& 12591&0513&08.45&0.16\cr
092246.94+230812.74&2319-53763-567&2.0&1.8 &06.71&20.33& 16993&0521&08.06&0.11\cr
094025.98+201707.30&2292-53713-048&2.0&2.8 &06.85&19.55& 13666&0819&07.63&0.14\cr
093059.16+202429.35&2289-53708-069&2.0&2.6 &07.45&19.90& 14229&0739&07.76&0.12\cr
091132.80+223200.66&2287-53705-169&2.0&1.9 &07.63&19.63& 11146&0210&08.30&0.14\cr
090554.64+213829.62&2284-53708-103&2.0&2.2 &07.37&19.52& 24597&0708&07.42&0.14\cr
083701.89+154454.66&2276-53712-004&2.0&1.9 &09.62&19.17& 15980&0329&08.16&0.08\cr
112439.29+262422.76&2216-53795-143&2.0& -  &06.48&19.48& 13047&0432&07.89&0.14\cr
140051.72+330754.37&2121-54180-369&2.0&1.9 &09.40&19.39& 22018&0548&07.81&0.09\cr
053507.04+001617.23&2072-53430-508&2.0&3.0 &10.11&19.96& 23486&0635&08.38&0.09\cr
110203.75+400558.02&1988-53469-434&2.0  &3.0 &05.79&19.80& 09511&0147&09.08&0.16 \cr
113804.16+305310.59&1974-53430-023&2.0  &2.0 &04.88&19.97& 11273&0373&09.08&0.27 \cr
155818.96+241758.75&1851-53524-476&2.0&2.0 &04.35&20.06& 13618&0850&09.13&0.20 \cr
155657.69+231358.47&1851-53524-298&2.0&2.2 &08.80&18.92& 18883&0538&07.36&0.10\cr
161658.43+070355.42&1731-53884-442&2.0&2.5 &07.17&19.45& 09415&0135&08.04&0.20 \cr
155651.96+351218.94&1417-53141-512&2.0&1.5 &08.59&19.54& 07582&0091&08.04&0.19 \cr
101529.62+090703.83&1237-52762-533&2.0&  K&12.24&18.60& 06903&0049&10.00&0.01\cr
153742.29+434719.76&1052-52466-619&2.0&  - &04.11&20.07& 37196&2306&07.67&0.36 \cr
145602.53-005548.75&0921-52380-528&2.0&2.0 &04.72&19.56& 14462&1133&08.08&0.21\cr
133420.98+041751.09&0853-52374-198&2.0&2.6 &12.65&18.53& 17415&0493&09.24&0.06\cr
155857.29+480047.44&0813-52354-328&2.0&2.6 &07.83&19.28& 14685&0556&09.02&0.12\cr
112022.02+635437.54&0597-52059-308&2.0&1.9 &05.19&19.27& 07561&0098&08.02&0.33\cr
144649.26+005215.44&0537-52027-126&2.0&2.2 &04.78&19.94& 13088&1128&08.33&0.23\cr
022523.68+002743.09&0406-51900-543&2.0&1.7 &05.47&20.01& 11661&1025&08.20&0.29\cr
171556.29+600643.89&0354-51792-318&2.0&  K&05.87&19.34& 12153&0797&08.46&0.29\cr
113213.00-003036.88&0282-51658-278&2.0&2.2 &07.44&19.82& 14286&0636&07.68&0.21\cr
152855.62-015148.79&0925-52411-312&1.8&  - &05.25&20.08& 21856&1156&07.64&0.23\cr
160904.12+175337.90&2967-54584-089&1.8&1.9 &08.61&19.20& 14379&0431&07.71&0.10\cr
084541.13+312936.60&2960-54561-009&1.8&2.0 &05.33&20.67& 11390&0283&08.93&0.22\cr
121033.24+221402.64&2644-54210-167&1.8&1.9 &34.91&17.07& 15144&0096&08.56&0.02\cr
\end{tabular}
\end{table}
\begin{table}
\begin{tabular}{lcrcrcrrrr}
Name (SDSS J)&P-M-F&$B_{H\alpha}$&$B_{H\beta}$&S/N&g&$T_\mathrm{eff}$&$\sigma_T$&log g&$\sigma_g$\cr
&&(MG)&(MG)&&(mag)&(K)&(K)&(cgs)&(cgs)\cr
131426.38+173228.14&2604-54484-481&1.8&1.9 &14.77&18.62& 07038&0054&09.95&0.06\cr
103648.67+171045.20&2593-54175-131&1.8&2.6 &06.22&19.63& 13884&1289&07.97&0.26\cr
092108.83+130111.65&2577-54086-448&1.8&2.1 &07.20&19.65& 13263&0538&07.84&0.13\cr
110911.12+582209.38&2414-54526-301&1.8&2.6 &06.74&20.30& 19207&0681&07.66&0.13\cr
100828.98+183633.49&2373-53768-290&1.8&2.0 &04.74&20.15& 09295&0177&08.74&0.23\cr
091310.44+230042.60&2287-53705-541&1.8&2.2 &06.42&20.03& 17629&0586&08.22&0.12\cr
081748.55+154341.15&2272-53713-386&1.8&1.2 &07.35&19.49& 17936&0460&08.21&0.11\cr
080703.25+135537.87&2268-53682-472&1.8&1.7 &06.79&19.88& 12887&0494&08.26&0.14\cr
163400.35+145651.38&2209-53907-472&1.8&2.2 &03.64&20.39& 13797&0820&08.06&0.22\cr
040054.82-064625.27&2071-53741-014&1.8&1.8 &08.56&19.64& 18869&0721&07.28&0.13\cr
004148.06-005127.47&1905-53706-253&1.8&2.6 &07.63&20.08& 13253&0551&08.07&0.13\cr
                   &0691-52199-300&3.3&3.9 &04.61\cr
233708.97+492532.00&1889-53240-584&1.8&2.2 &09.97&19.58& 09859&0100&08.23&0.12\cr
123527.09+145318.64&1768-53442-074&1.8&2.2 &06.55&19.55& 15250&0466&07.78&0.11\cr
232937.55+524437.90&1663-52973-119&1.8&1.8 &14.00&18.95& 32204&0370&08.22&0.08\cr
080150.50+205012.05&1583-52941-599&1.8&  - &03.51&20.43& 11831&1244&08.32&0.35 \cr
214539.84+000136.44&0990-52465-080&1.8&  - &07.16&19.55& 30307&0802&07.92&0.18 \cr
010654.65-104315.01&0659-52199-207&1.8&1.7 &09.57&19.19& 08573&0102&07.70&0.22\cr
094103.85+032835.84&0570-52266-279&1.8&2.2 &04.69&20.07& 19333&0964&07.88&0.27\cr
113431.97-031528.99&0327-52294-131&1.8&1.7 &07.09&19.57& 16750&0584&09.12&0.10\cr
144114.21+003702.40&0307-51663-595&1.8&1.8 &06.57&19.87& 16750&0507&08.07&0.11\cr
075036.55+222021.48&2916-54507-133&1.7&1.7 &08.18&20.29& 18254&0555&07.82&0.11\cr
035010.32+085829.18&2697-54389-048&1.7&1.7 &06.47&20.32& 11462&0558&07.81&0.23\cr
071814.18+305148.75&2695-54409-268&1.7&1.7 &05.88&20.47& 12248&0839&08.12&0.21\cr
064607.86+280510.14&2694-54199-175&1.7&2.1 &20.67&18.69& 22775&0288&09.05&0.04\cr
172623.14+632607.81&2561-54597-212&1.7&1.6 &07.16&20.42& 18182&0643&08.07&0.13\cr
163013.93+123941.95&2533-54585-325&1.7&2.2 &07.17&19.86& 12124&0289&08.11&0.13\cr
160532.19+131748.69&2525-54569-425&1.7&1.9 &04.00&20.28& 12402&0549&07.65&0.22\cr
154424.84+132650.84&2517-54567-110&1.7&1.7 &11.80&18.97& 13793&0736&07.93&0.09\cr
030158.92+372204.26&2443-54082-190&1.7& -  &12.67&19.57& 28949&0432&08.06&0.09\cr
084522.95+150020.36&2429-53799-152&1.7&2.3 &08.36&19.35& 17904&0432&07.79&0.10\cr
084039.37+125706.28&2428-53801-354&1.7&1.7 &05.39&19.63& 10604&0232&07.97&0.20\cr
082447.49+131543.26&2422-54096-588&1.7&1.8 &05.91&19.45& 19134&0545&07.81&0.14\cr
104851.94+273817.72&2358-53797-156&1.7&1.9 &07.33&19.20& 10923&0175&07.75&0.14\cr
021338.52+053023.40&2321-53711-125&1.7&2.2 &06.27&20.20& 09058&0132&08.14&0.23\cr
053400.84+625419.80&2302-53709-579&1.7&2.1 &05.24&20.81& 10954&0569&08.11&0.23\cr
091002.83+232220.03&2287-53705-470&1.7&1.7 &06.17&19.94& 15291&0538&07.76&0.12\cr
084845.66+214047.05&2280-53680-446&1.7&1.7 &07.74&19.73& 19142&0502&07.86&0.10\cr
224602.82+230704.18&2261-53612-559&1.7&1.8 &10.63&19.56& 17160&0327&08.07&0.07\cr
113500.53+291206.62&2220-53795-222&1.7& -  &06.31&19.65& 21456&0667&07.79&0.14\cr
161321.41+155332.09&2198-53918-400&1.7&1.7 &05.01&19.92& 09959&0189&07.98&0.25\cr
174755.72+251232.33&2194-53904-154&1.7& -  &09.34&19.75& 10390&0122&08.32&0.14\cr
074327.09+273732.37&2075-53730-278&1.7&1.5 &11.21&19.88& 08727&0101&07.07&0.36 \cr
074047.65+180907.32&2074-53437-237&1.7&  - &14.31&19.27& 17904&0299&07.94&0.06\cr
073135.37+353108.57&2073-53728-108&1.7&1.5 &05.30&20.36& 22795&1052&07.85&0.17 \cr
052831.03+005244.22&2072-53430-336&1.7&1.5 &14.13&19.52& 16444&0300&08.03&0.06\cr
053016.83-001034.50&2072-53430-233&1.7&1.7 &05.79&20.44& 16128&0888&07.87&0.17\cr
084253.04+092226.54&1759-53081-618&1.7&2.1 &09.14&19.15& 11316&0955&07.91&0.17\cr
083002.71+083632.04&1758-53084-531&1.7&1.6 &08.21&19.42& 15642&0388&07.81&0.09\cr
134758.89+495427.49&1669-53433-020&1.7&1.9 &19.25&17.79& 16628&0181&08.56&0.04\cr
144541.72+411441.59&1397-53119-352&1.7&1.0 &06.06&19.94& 24938&0802&07.25&0.20 \cr
014938.34-004938.03&0699-52202-012&1.7&2.6 &07.23&19.35& 16010&0488&07.89&0.11\cr
082159.66+431127.28&0547-52207-019&1.7& -  &05.36&20.15& 09646&0901&05.12&0.13\cr
162304.11+183522.23&2969-54586-617&1.5&1.6 &10.64&19.16& 14169&0665&08.08&0.08\cr
075704.00+085520.06&2945-54505-148&1.5&1.2 &09.92&20.01& 18123&0424&07.95&0.09\cr
044512.39-052524.52&2942-54521-487&1.5& -  &06.52&20.44& 12488&0430&08.62&0.15\cr
192553.60+620708.70&2563-54653-165&1.5&1.7 &06.54&19.94& 10504&0177&09.04&0.18\cr
030550.35+370759.16&2443-54082-154&1.5& -  &10.04&19.79& 12414&0483&07.93&0.13\cr
084111.34+154921.03&2429-53799-358&1.5&1.3 &06.67&19.79& 18090&0487&07.91&0.12\cr
004011.49+070255.73&2327-53710-074&1.5&1.7 &12.11&19.52& 18529&0381&08.09&0.07\cr
212143.08-060005.77&2320-54653-445&1.5&  - &12.02&19.74& 25075&0624&08.02&0.08\cr
\end{tabular}
\end{table}

\begin{table}
\begin{tabular}{lcrcrcrrrr}
Name (SDSS J)&P-M-F&$B_{H\alpha}$&$B_{H\beta}$&S/N&g&$T_\mathrm{eff}$&$\sigma_T$&log g&$\sigma_g$\cr
&&(MG)&(MG)&&(mag)&(K)&(K)&(cgs)&(cgs)\cr
083020.36+185814.55&2275-53709-229&1.5& -  &12.78&19.02& 22490&0432&08.69&0.07\cr
203332.94+140115.37&2258-54328-295&1.5&1.3 &11.02&20.22& 18529&0436&07.78&0.08\cr
235107.48+403454.05&1883-53271-521&1.5&2.2 &07.67&20.14& 16535&0505&08.92&0.10\cr
080719.89+064536.58&1756-53080-234&1.5&1.5 &04.49&20.16& 12380&0912&07.92&0.29 \cr
162115.35+075059.05&1732-53501-455&1.5&1.7 &06.72&19.83& 16750&0501&07.75&0.12\cr
141808.13+481850.59&1672-53460-181&1.5&2.6 &08.95&19.10& 14310&0313&06.99&0.10\cr
203016.13+765022.67&1661-53240-116&1.5&2.0 &05.11&20.31& 13277&1567&07.72&0.26 \cr
143218.26+430126.72&1396-53112-338&1.5&  - &09.83&19.00& 25429&0590&07.94&0.10\cr
112328.49+095619.39&1222-52763-625&1.5&1.7 &18.25&17.74& 10537&0057&08.70&0.06\cr
160929.97+443857.20&0814-52370-317&1.5&1.7 &05.40&19.75& 40320&2355&08.17&0.32\cr
145029.51+032218.84&0587-52026-016&1.5& -  &09.45&19.12& 14261&0444&07.85&0.09\cr
095603.30+540907.05&0769-54530-078&1.4&1.7 &07.88&19.51& 12180&0456&07.58&0.17\cr
154120.14+022756.54&0594-52045-315&1.4&1.7 &05.33&20.05& 19937&0923&07.66&0.21\cr
065133.34+284423.44&2694-54199-528&1.3&1.6 &18.18&19.11& 17841&0265&07.06&0.05\cr
122238.87+005034.42&2568-54153-411&1.3&1.7 &11.39&19.87& 12956&0470&07.86&0.09\cr
163630.30+114452.39&2533-54585-502&1.3&1.6 &05.77&20.30& 10582&0215&08.48&0.20\cr
093813.84+615600.92&2403-53795-278&1.3&1.6 &08.63&20.14& 14004&0475&07.80&0.10\cr
101642.27+281610.22&2348-53757-040&1.3&1.7 &05.46&19.81& 24231&0897&07.80&0.17\cr
212232.58-061839.74&2320-54653-537&1.3&0   &17.06&19.38& 26168&0557&07.70&0.07\cr
130535.77+283014.60&2242-54153-447&1.3& -  &09.68&19.36& 28932&0465&07.90&0.10\cr
082107.35+194433.68&2082-53358-617&1.3& -  &08.84&19.13& 25636&0795&07.44&0.11\cr
004038.90+243852.77&2058-53349-195&1.3& -  &12.05&19.72& 14375&0385&07.79&0.08\cr
105152.19+321135.19&2026-53711-183&1.3&2.0 &05.74&20.01& 14529&0741&09.08&0.16 \cr
123449.89+150348.87&1768-53442-557&1.3&  - &13.20&18.81& 06302&0007&10.00&0.01 \cr
155932.58+081904.62&1728-53228-175&1.3&1.0 &04.81&20.22& 10080&0206&08.26&0.28 \cr
133007.57+104830.59&1699-53148-137&1.3&2.0 &11.52&18.90& 07094&0065&08.56&0.13 \cr
170657.90+232118.99&1687-53260-019&1.3&1.7 &07.79&19.24& 22242&0648&08.07&0.10\cr
201822.90+754807.62&1661-53240-124&1.3&2.5 &05.10&20.32& 14694&3140&07.94&0.24\cr
203828.49+764123.07&1661-53240-023&1.3&2.0 &06.03&19.90& 11138&1504&08.82&0.27 \cr
161118.60+242446.65&1657-53520-372&1.3&1.5 &09.93&19.13& 08935&0082&07.85&0.13 \cr
142118.18+523547.17&1326-52764-250&1.3&1.5 &04.57&19.71& 21752&0836&08.47&0.17 \cr
083745.13+064313.82&1297-52963-637&1.3& -  &07.41&19.74& 25966&0829&07.91&0.12\cr
093411.36+364641.28&1275-52996-174&1.3&2.0 &06.56&19.68& 13930&0604&09.03&0.14 \cr
084628.05+064532.83&1189-52668-344&1.3&2.0 &04.23&20.05& 17143&0781&08.27&0.24 \cr
161030.50+365442.22&1056-52764-440&1.3&2.0 &08.94&19.24& 19734&0448&09.28&0.08 \cr
100932.74+524638.21&0903-52400-584&1.3&2.0 &16.11&18.60& 15122&0295&09.09&0.05\cr
093508.57+042116.76&0569-52264-452&1.3& -  &06.90&19.86& 11468&0697&08.17&0.20\cr
092932.56+561318.58&0556-51991-326&1.3&2.0 &07.93&19.61& 08935&0093&07.77&0.18\cr
121557.43+665344.84&0493-51957-104&1.3& -  &09.78&19.14& 06935&0087&09.18&0.29\cr
031630.64-081529.03&0459-51924-002&1.3&1.6 &11.01&19.05& 11677&0203&09.09&0.11\cr
075916.54+433519.06&0436-51883-045&1.3&2.1 &11.63&18.74& 22420&0459&09.62&0.08\cr
012641.92+132537.21&0425-51844-283&1.3& -  &11.40&18.60& 08566&0052&05.04&0.05\cr
022623.81-002313.09&0406-52238-071&1.3&1.7 &05.13&19.80& 08727&0134&05.07&0.08\cr
022335.16+004954.90&0406-51900-490&1.3& -  &07.16&19.77& 07501&0097&07.69&0.21\cr
220514.08-005841.67&0373-51788-243&1.3& -  &11.87&18.69& 32204&0473&07.83&0.11\cr
121706.47+172856.02&2596-54207-023&1.2&  - &07.07&19.16& 23729&0838&08.09&0.14\cr
083051.08+244615.73&2330-53738-503&1.2&2.2 &10.85&19.78& 16454&0355&08.02&0.07\cr
013742.54+235138.27&2064-53341-491&1.2&1.7 &07.14&20.24& 14043&4018&08.37&0.39\cr
233039.04+500729.68&1889-53240-377&1.2&  - &09.86&19.53& 10811&0139&08.29&0.10\cr
082239.43+082436.75&1758-53084-346&1.2&  - &18.83&18.12& 11193&0077&08.56&0.05\cr
165354.58+251738.46&1692-53473-213&1.2&1.0 &03.78&20.35& 12556&1473&07.88&0.30 \cr
172705.00+084857.15&2818-54616-374&1.0&  - &28.38&18.42& 41618&0550&08.02&0.06\cr
140444.22+201922.62&2771-54527-196&1.0&1.3 &10.31&19.15& 13472&0367&08.12&0.08\cr
093431.11+132814.70&2580-54092-273&1.0&  - &10.85&19.02& 34332&0631&08.07&0.12\cr
084233.37+101806.35&2573-54061-141&1.0&  - &11.20&19.06& 12284&0225&08.40&0.09\cr
155232.77+264636.67&2474-54564-392&1.0&  - &11.18&19.89& 19362&0409&07.89&0.08\cr
124508.48+591551.78&2461-54570-264&1.0& -  &07.93&19.94& 17184&0475&08.04&0.10\cr
133836.07+652433.09&2460-54616-042&1.0&  - &09.55&19.73& 22733&0667&07.93&0.10\cr
024241.67+291608.32&2444-54082-604&1.0&1.1 &12.63&19.44& 29405&0429&07.80&0.08\cr
\end{tabular}
\end{table}
\begin{table}
\begin{tabular}{lcrcrcrrrr}
Name (SDSS J)&P-M-F&$B_{H\alpha}$&$B_{H\beta}$&S/N&g&$T_\mathrm{eff}$&$\sigma_T$&log g&$\sigma_g$\cr
&&(MG)&(MG)&&(mag)&(K)&(K)&(cgs)&(cgs)\cr
103403.99+305034.45&2354-53799-536&1.0&  - &06.38&19.51& 14876&0430&07.89&0.11\cr
212514.19-062152.52&2320-54653-612&1.0&  - &07.72&20.25& 11242&0246&08.39&0.15\cr
211744.75-073652.87&2320-54653-312&1.0&  - &12.54&19.53& 32642&0486&09.51&0.10\cr
093654.95+262650.24&2294-54524-617&1.0&1.6 &08.25&19.61& 15691&0401&07.92&0.09\cr
091326.64+211250.50&2288-53699-332&1.0& -  &09.65&18.82& 14512&0430&08.14&0.08\cr
131702.34+281848.62&2243-53794-533&1.0& -  &10.90&19.08& 11481&0175&08.32&0.09\cr
172735.81+280536.92&2193-53888-570&1.0& -  &14.64&19.37& 18084&0292&07.91&0.06\cr
154605.41+243759.05&1850-53786-312&1.0&  - &08.82&18.78& 06438&0067&09.29&0.13 \cr
140822.00+443007.96&1467-53115-557&1.0&  - &10.49&19.02& 15859&0279&07.21&0.08\cr
160246.46+303914.48&1405-52826-283&1.0& -  &37.79&16.06& 60359&1229&07.32&0.07\cr
060442.49+641357.12&2301-53712-476&$\leq$ 1 & -  &19.01&19.09& 67833&3013&10.00&0.00\cr
132926.05+254936.50&2245-54208-307&$\leq$ 1 & -  &23.52&17.42& 28299&0175&09.39&0.04\cr
080527.56+073534.16&2056-53463-557&$\leq$ 1 &  - &24.66&17.89& 06300&0003&10.00&0.01 \cr
165538.93+253345.99&1692-53473-163&$\leq$ 1 & -  &29.73&16.94& 11141&0052&09.38&0.03\cr
081018.73+370010.95&0892-52378-374&$\leq$ 1 & -  &21.31&18.14& 09556&0042&08.37&0.05\cr
143632.86+563525.43&0791-52435-304&$\leq$ 1 &  - &07.50&18.70& 06865&0079&09.85&0.14\cr
075842.68+365731.59&0757-52238-439&$\leq$ 1 &  - &14.61&19.01& 06760&0048&08.15&0.11 \cr
162727.70+492507.99&0625-52145-578&$\leq$ 1 & -  &10.33&18.98& 08449&0079&09.96&0.06\cr
145636.33+583321.24&0610-52056-189&$\leq$ 1 &  - &06.30&19.64& 07601&0100&07.85&0.30 \cr
100732.64+010914.60&0501-52235-004&$\leq$ 1 & -  &07.29&19.58& 07215&0101&07.39&0.27 \cr
085309.10+563441.51&0483-51924-203&$\leq$ 1 &-   &07.92&19.62& 07889&0101&07.77&0.22\cr
173235.20+590533.46&0366-52017-591&$\leq$ 1 & -  &12.92&18.73& 10818&0098&08.03&0.09\cr
\end{tabular}
\caption{Magnetic White Dwarf Stars. Notes: 
P-M-F are the Plate-Modified Julian Date-Fiber number that designates an
SDSS spectrum.
K in the H$\beta$ magnetic field column means it was measured by \citet{Kulebi}. For those spectra we quote only their field determinations, as they fit the whole spectrum, keeping $\log g=8.0$.*SDSS J125044.42+154957.3 is a $P_\mathrm{orb}=86$m binary \citep{Breedt12}.}
\end{table}
\label{lastpage}
\end{document}